\documentclass[journal,11pt,onecolumn]{IEEEtran}
\usepackage{amsmath,amssymb,epsfig,psfrag,cite,subfigure}
\include{macros}
\usepackage{graphicx}

\usepackage{algorithm}
\usepackage{epsfig,psfrag}
\usepackage{subfigure}
\usepackage{color}
\usepackage{url}

\newcommand{\bs}[1]{\boldsymbol{#1}}

\newcommand{\CRLB}{{\rm{CRLB}}}

\newcommand{\SNR}{{\rm{SNR}}}

\newcommand{\SNRt}{{\rm{SNR}_{\rm{thr}}}}

\renewcommand{\baselinestretch}{1.7}

\begin{document}

\title{Optimal Jammer Placement in Wireless Localization Systems}

\author{Sinan Gezici,\thanks{Copyright (c) 2016 IEEE. Personal use of this material is permitted. However, permission to use this material for any other purposes must be obtained from the IEEE by sending a request to pubs-permissions@ieee.org.} Suat Bayram,\thanks{S. Gezici and M. N. Kurt are with the Department of Electrical and Electronics Engineering, Bilkent University, 06800, Ankara, Turkey, Tel: +90-312-290-3139, Fax: +90-312-266-4192, Emails: \{gezici,mnkurt\}@ee.bilkent.edu.tr. S. Bayram is with the Department of Electrical and Electronics Engineering, Turgut Ozal University, 06010, Ankara, Turkey, Email: sbayram@turgutozal.edu.tr. M. R. Gholami is with ACCESS Linnaeus Center, Electrical Engineering, KTH Royal Institute of Technology, 100 44, Stockholm, Sweden, Email: mohrg@kth.se.} Mehmet Necip Kurt, and Mohammad Reza Gholami\thanks{This work was supported in part by the Distinguished Young Scientist Award of Turkish Academy of Sciences (TUBA-GEBIP 2013). The authors would like to thank Dr. Magnus Jansson for his useful comments. Part of this work was presented at IEEE 16th International Workshop on Signal Processing Advances in Wireless Communications (SPAWC), 2015, Stockholm, Sweden, June 2015 \cite{GeziciSPAWC15}.}\vspace{-0.5cm}}

\maketitle

\begin{abstract}
  In this study, the optimal jammer placement problem is proposed and analyzed for wireless localization systems. In particular, the optimal location of a jammer node is obtained by maximizing the minimum of the Cram{\'e}r-Rao lower bounds (CRLBs) for a number of target nodes under location related constraints for the jammer node. For scenarios with more than two target nodes, theoretical results are derived to specify conditions under which the jammer node is located as close to a certain target node as possible, or the optimal location of the jammer node is determined by two of the target nodes. Also, explicit expressions are provided for the optimal location of the jammer node in the presence of two target nodes. In addition, in the absence of distance constraints for the jammer node, it is proved, for scenarios with more than two target nodes, that the optimal jammer location lies on the convex hull formed by the locations of the target nodes and is determined by two or three of the target nodes, which have equalized CRLBs. Numerical examples are presented to provide illustrations of the theoretical results in different scenarios.

\end{abstract}

\vspace{-0.2cm}

\section{Introduction}\label{sec:Intro}

Position information has a critical role for various location aware applications and services in current and next generation wireless systems \cite{Sinan_book,HandbookPos}. In the absence of GPS signals, e.g., due to lack of access to GPS satellites in some indoor environments, position information is commonly extracted from a network consisting of a number of anchor nodes at known locations via measurements of position related parameters such as time-of-arrival (TOA) or received signal strength (RSS) \cite{HandbookPos}. In such wireless localization networks, the aim is to achieve high localization accuracy, which is commonly defined in terms of the mean squared position error \cite{Sinan_Survey}.

Jamming can degrade performance of wireless localization systems and can have significant effects in certain scenarios.
Although jamming and anti-jamming approaches are investigated for GPS systems in various studies such as \cite{GPSjam,GPSantiJam_CLEAN,GPSanti2}, effects of jamming on wireless localization networks have gathered little attention in the literature. Recently, a wireless localization network is investigated in the presence of jammer nodes, which aim to degrade the localization accuracy of the network, and the optimal power allocation strategies are proposed for the jammer nodes to maximize the average or the minimum Cram\'{e}r-Rao lower bounds (CRLBs) of the target nodes \cite{Gezici_JammerPowAlloc1}. The results provide guidelines for quantifying the effects of jamming in wireless localization systems \cite{Gezici_JammerPowAlloc1}.

The study in \cite{Gezici_JammerPowAlloc1} assumes fixed locations for the jammer nodes and aims to perform optimal power allocation{, which leads to convex (linear) optimization problems}. In this manuscript, the main purpose is to determine the optimal location of a jammer node in order to achieve the best jamming performance in a wireless localization network consisting of multiple target nodes. In particular, the optimal location of the jammer node is investigated to maximize the minimum of the CRLBs for the target nodes in a wireless localization network in the presence of constraints on the location of the jammer node. Although there exist some studies that investigate the jammer placement problem for communication systems, e.g., to prevent eavesdroppers \cite{ProtectiveJamming} or to jam wireless mesh networks \cite{a2}, the optimal jammer placement problem has not been considered for wireless localization networks in the literature (see \cite{GeziciSPAWC15} for the conference version of this study).

\subsection{Literature Survey on Node Placement}

Optimal node placement has been studied intensely for wireless sensor networks (WSNs) in the last decade, and various objectives have been considered for placement of sensor nodes. For example, in \cite{a9} and \cite{a6}, the aim is to provide complete coverage of the WSN area with the minimum number of sensor nodes. In \cite{a7}, the aim is to maximize the lifetime of the network via distance based placement whereas the resilience of the network to single node failures is the main objective in \cite{a10}. In another study, powerful relay nodes are placed together with sensor nodes in order to increase the lifetime of the network \cite{a8}.

Placement of jammer nodes in wireless networks can be performed for various purposes \cite{a11}.
While the aim of jammer placement is generally to create disruptive effects on the network operation, different objectives are also considered in the literature. In \cite{a1}, the aim is to divide network into subparts and to prevent the network traffic between those subparts via jamming. In \cite{a2}, the main objective is to destroy the communication links in the network in the worst possible way by placing jammer nodes efficiently. On the other hand, in \cite{ProtectiveJamming}, the purpose of using jammer nodes is to protect the network from eavesdroppers, and the function of jammer nodes is to reduce signal quality below a level such that no illegitimate receiver can reach the network data. During this protection, signal quality must be kept above a certain level for other devices so that the actual network operation is not prevented. Based on these two main criteria, the optimal placement of jammer nodes are performed in \cite{ProtectiveJamming}.

Against jamming attacks, various anti-jamming techniques have also been developed \cite{a12,a14,a16,a17,a19,a20,a22}. Some studies such as \cite{a17} focus on finding positions of jamming devices for taking security actions against them; e.g., physically destroying them or changing the routing protocol, in order not to traverse the jammed region \cite{a17}. Another technique is to rearrange the positions of the nodes in the network after each attack in order to mitigate the effects of jamming \cite{a22}. In addition, \cite{a11} employs a game theoretic approach, in which the attacker tries to maximize the damage on the network activity while the aim of the defender is to secure the multi-hop multi-channel network. Actions available to the attacker are related to choosing the positions of jammer nodes and the channel hopping strategy while the action of the defender is based on choosing the channel hopping strategy.

In the literature, there also exist some practical heuristic approaches for node placement. In case of jamming, placing jammer nodes close to source and destination nodes, at the critical transshipment points of the network, or where sensor nodes are dense are among such approaches \cite{a2}. By evaluating efficiency of different jammer locations, these heuristic approaches can be analyzed and compared for various scenarios. In some studies such as \cite{ProtectiveJamming}, the best jammer location is chosen among finitely many predetermined locations. The motivation behind this method is that it is not always possible to place jammer nodes at desired locations due to topological limitations, risk of visual detection by enemies, or tight security measurements \cite{a2}. In addition, for both jammer and sensor node placement, the grid base approach is widely employed. In this approach, the continuous sensor field is divided into equal-area grid cells and the best location is determined via evaluation over finite set of points. As the grid size is reduced, performance of node placement improves in general; however, the required computational effort to find the best location increases as well. In \cite{a2}, based on the grid-based approach, it is shown that the most disruptive effect on the network occurs when jammer nodes are placed close to source and destination nodes. Similarly, in \cite{a11}, it is stated that
the optimal solution for jammer nodes is to jam the network flow concentrated near source and destination nodes.

Placement of anchor nodes has been studied for wireless localization systems, in which the aim is to perform optimal deployment of anchor nodes for improving localization accuracy of target nodes in the system \cite{a23,a26,a24,a27}.
For example, in \cite{a26}, placement of anchor nodes is performed in order to minimize the CRLB in an RSS based localization system. On the other hand, the authors in \cite{a27} employ an optimization method based on integer-coded genetic algorithm for optimizing the average localization error and the signal coverage estimate.


\subsection{Contributions}

Although placement of anchor nodes is considered for wireless localization systems (e.g., \cite{a23,a24,a26,a27}) and placement of jammer nodes is studied for communication systems (e.g., \cite{a1,a2,ProtectiveJamming}), there exist no studies that investigate the problem of optimal jammer placement in wireless localization systems.
In this manuscript, the optimal jammer placement problem is proposed and analyzed for wireless localization systems. In particular, the minimum of the CRLBs of the target nodes is
considered as the objective function (to guarantee that all the target nodes have localization accuracy bounded by a certain limit) and constraints are imposed on distances between the jammer node and target nodes. In addition to the generic formulation, which leads to a non-convex problem, various special cases are investigated and theoretical results are presented to characterize the optimal solution. Especially, the scenario with two target nodes and the scenario {with more than two target nodes and} in the absence of distance constraints are investigated in detail. Various numerical examples are presented to verify and explain the theoretical results. The main contributions of this manuscript can be summarized as follows:
\begin{itemize}
\item The optimal jammer placement problem in a wireless localization system is proposed for the first time.
\item {In the presence of more than two target nodes,} conditions are derived to specify scenarios in which the optimal jammer location is as close to a certain target node as possible (Proposition~1) or the jammer node is located on the straight line that connects two target nodes (Proposition~2). In addition, for the case of two target nodes, the optimal location of the jammer node is specified explicitly (Proposition~3).
\item In the absence of distance constraints for the jammer node, it is proved{, for scenarios with more than two target nodes,} that the optimal location of the jammer node lies on the convex hull formed by the locations of the target nodes (Proposition~4), {where the projection theorem is utilized for specifying the location of the jammer node.}
\item For scenarios with three target nodes and in the absence of distance constraints, it is shown that the optimal jammer location equalizes the CRLBs of either all the target nodes or two of the target nodes, which correspond to cases in which the jammer node lies on the interior or on the boundary of the triangle formed by the target nodes, respectively (Propositions~5 and 6-(a)). In addition, a necessary and sufficient condition is presented for the optimal jammer location to be on the interior or the boundary of that triangle (Proposition~6-(b)).
\item In the absence of distance constraints for the jammer node {and in the presence of more than three target nodes}, it is proved that the optimal jammer location is determined by two or three of the target nodes (Proposition~7).
\end{itemize}
The main motivations behind the study of the optimal jammer placement problem for wireless localization are related to performing efficient jamming of a wireless localization system (e.g., of an enemy) to degrade localization accuracy, and presenting theoretical results on optimal jamming performance, which can be useful for providing guidelines for developing anti-jamming techniques {(see Section~\ref{sec:Conc}).


\section{System Model}\label{sec:SysModel}

Consider a wireless localization network in a two-dimensional space consisting of $N_A$ anchor nodes and $N_T$ target nodes located at $\bs{y}_i\in\mathbb{R}^2$, $i=1,\ldots,N_A$ and $\bs{x}_i\in\mathbb{R}^2$, $i=1,\ldots, N_T$, respectively. It is assumed that $\bs{x}_i$'s ($\bs{y}_i$'s) are all distinct. The target nodes are assumed to estimate their locations based on received signals from the anchor nodes, which have known locations; i.e., self-positioning is considered \cite{Sinan_Survey}.
In addition to the target and anchor nodes, there exists a jammer node at location $\bs{z}\in\mathbb{R}^2$, which aims to degrade the localization performance of the network. The jammer node is assumed to transmit zero-mean Gaussian noise, as commonly employed in the literature \cite{a2,SimonBook,WeissJam,McJam}.

In this manuscript, non-cooperative localization is studied, where target nodes receive signals only from anchor nodes (i.e., not from other target nodes) for localization purposes. Also, the connectivity sets are defined as $\mathcal{A}_i\triangleq \{j\in\{1,\ldots,N_A\}~|~\text{anchor node}~j~ \text{is connected to target node}~ i \}$ for $i\in\{1,\ldots,N_T\}$. Then, the received signal at target node $i$ coming from anchor node $j$ is expressed as \cite{Gezici_JammerPowAlloc1}
\begin{align}\label{eq:model}
{r_{ij}(t)=\sum_{k=1}^{L_{ij}}\alpha^k_{ij}s_j(t-\tau_{ij}^k)
+\gamma_{ij}\sqrt{P_J}\,v_{ij}(t)
+n_{ij}(t)}
\end{align}
for $t\in[0,T_{\rm{obs}}]$, $i\in\{1,\ldots,N_T\}$, and $j\in\mathcal{A}_i$, where $T_{\rm{obs}}$ is the observation time, $\alpha^k_{ij}$ and $\tau_{ij}^k$ represent, respectively, the amplitude and delay of the $k$th multipath component between anchor node $j$ and target node $i$, $L_{ij}$ is the number of paths between target node $i$ and anchor node $j$, {$P_J$ is the transmit power of the jammer node, and $\gamma_{ij}$ denotes the channel coefficient between target node~$i$ and the jammer node during the reception of the signal from anchor node~$j$.} The transmit signal $s_{{j}}(t)$ is known, and the measurement noise $n_{ij}(t)$ and the jammer noise $\sqrt{P_J}\,{v_{ij}(t)}$ are assumed to be independent zero-mean white Gaussian random processes\footnote{Even though it is theoretically possible to mitigate the effects of zero-mean white Gaussian noise by repeating measurements, the observation interval (the number of measurements) cannot be increased arbitrarily in practical localization systems since the location of a target node should approximately be constant during the observation interval. Also, increasing the observation interval for localization can lead to data rate reduction in systems that perform both localization and data transmission. {When multiple independent measurements are taken, the $\lambda_{ij}$ term in \eqref{eq:lambda_i} can be scaled by the number of measurements.}}, where the spectral density level of $n_{ij}(t)$ is $N_0/2$ and that of ${v_{ij}(t)}$ is equal to one \cite{Gezici_JammerPowAlloc1}. {Also, for each $i\in\{1,\ldots,N_T\}$, $n_{ij}(t)$'s ($v_{{ij}}(t)$'s) are assumed to be independent for $j\in\mathcal{A}_i$.}\footnote{{The transmitted signals, $s_j(t)$'s, are assumed to be orthogonal \cite{Win_2014_IEEEnetw_PowerOpt} (cf. Remark~1).}}
The delay $\tau_{ij}^k$ is expressed as
\begin{gather}
\tau_{ij}^k = \left(\|\bs{y}_j-\bs{x}_i\|+b_{ij}^k\right)/\,c
\end{gather}
with $b_{ij}^k\geq 0$ representing a range bias and $c$ being the speed of propagation. Set $\mathcal{A}_i$ is partitioned as follows: $\mathcal{A}_i\triangleq \mathcal{A}_i^{L}\cup\mathcal{A}_i^{NL}$,
where $\mathcal{A}_i^{L}$ and $\mathcal{A}_i^{NL}$ denote the sets of anchors nodes with line-of-sight (LOS) and non-line-of-sight (NLOS) connections to target node $i$, respectively.

{It is noted from \eqref{eq:model} that a constant jamming attack is considered in this study, where the jammer node constantly emits white Gaussian noise \cite{WorstCaseJammingPoor,ISITjam07}. This model is well-suited for scenarios in which the jammer node has the ability to transmit noise only, or does not know the ranging signals employed between the anchor and target nodes. In such scenarios, the jammer node can constantly transmit Gaussian noise for efficient jamming as the Gaussian distribution corresponds to the worst-case scenario among all possible noise distributions according to some criteria such as minimizing the mutual information and maximizing the mean-squared error \cite{WorstCaseJam,WorstAdditive,Kay_93}.}

{\textit{\textbf{Remark 1:}} In practical wireless localization systems, multiple access  techniques, such as time division multiple access or frequency division multiple access, are employed so that the signal from each anchor node can be observed by each target node without any interference from the other anchor nodes, as stated in \eqref{eq:model} \cite{Win_2014_IEEEnetw_PowerOpt}. Therefore, for each target node, the received signals related to different anchor nodes contain jamming signals that correspond to different time intervals or frequency bands; hence, for each $i$, $v_{ij}(t)$ for $j\in\mathcal{A}_i$ can be modeled as independent.}

\section{CRLBs for Localization of Target Nodes}\label{sec:CRLB}

Regarding target node $i$, the following vector consisting of the bias terms in the LOS and NLOS cases is defined \cite{Yihong_VTC}:
\begin{gather}\label{eq:biasTerms}
\bs{b}_{ij}=
\begin{cases}
\left[b_{ij}^2 \ldots b_{ij}^{L_{ij}}\right]^T\,,&\text{if}~j\in\mathcal{A}_i^L\\
\left[b_{ij}^1 \ldots b_{ij}^{L_{ij}}\right]^T\,,&\text{if}~j\in\mathcal{A}_i^{NL}
\end{cases}.
\end{gather}
From \eqref{eq:biasTerms}, the unknown parameters related to target node $i$ are defined as follows \cite{shen2010fundamental}:
\begin{align}
\bs{\theta}_i\triangleq
\left[\bs{x}_i^T~\bs{b}^T_{i\mathcal{A}_i(1)}~\cdots~\bs{b}^T_{i{\mathcal{A}_i(|\mathcal{A}_i|)}}
~\bs{\alpha}^T_{i\mathcal{A}_i(1)}~\cdots~\bs{\alpha}^T_{i{\mathcal{A}_i(|\mathcal{A}_i|)}}\right]^T
\end{align}
where $\mathcal{A}_i(j)$ denotes the $j$th element of set $\mathcal{A}_i$, $|\mathcal{A}_i|$ represents the number of elements in $\mathcal{A}_i$, and $\bs{\alpha}_{ij}=\big[\alpha_{ij}^1\cdots\alpha_{ij}^{L_{ij}}\big]^T$.
The total noise level is assumed to be known by each target node.

The CRLB for location estimation is expressed as~\cite{shen2010fundamental}
\begin{align}\label{eq:CRLBineq}
\mathbb{E}\left\{\|\hat{\bs{x}}_i-\bs{x}_i\|^2\right\}\geq {\rm{tr}}\left\{\left[\bs{F}_i^{-1}\right]_{2\times2}\right\}
\end{align}
where $\hat{\bs{x}}_i$ represents an unbiased estimate of the location of target node $i$, $\rm{tr}$ denotes the trace operator, and $\bs{F}_i$ is the Fisher information matrix for vector $\bs{\theta}_i$. Based on the steps in~\cite{shen2010fundamental},  $\left[\bs{F}_i^{-1}\right]_{2\times2}$ in \eqref{eq:CRLBineq} can be stated as
\begin{gather}\label{eq:CRLB1}
\left[\bs{F}_i^{-1}\right]_{2\times2}=\bs{J}_i(\bs{x}_i,P_J)^{-1}
\end{gather}
where the equivalent Fisher information matrix $\bs{J}_i(\bs{x}_i,P_J)$ in the absence of prior information about the location of the target node is expressed as {(see Theorem~1 in \cite{shen2010fundamental} for the derivations)}
\begin{align}\label{eq:fisher}
\bs{J}_i(\bs{x}_i,P_J)=\sum_{j\in\mathcal{A}_i^L}\frac{\lambda_{ij}}{N_0/2
+P_J|\gamma_{i{j}}|^2}\,\bs{\phi}_{ij}\bs{\phi}^T_{ij}
\end{align}
with
\begin{align}\label{eq:lambda_i}
\lambda_{ij}&\triangleq\frac{4\pi^2\beta_{{j}}^2|\alpha_{ij}^1|^2\int_{-\infty}^{\infty}
|S_{{j}}(f)|^2df}{c^2}(1-\xi_{ij})\,,
\\\label{eq:phiVec}
\bs{\phi}_{ij}&\triangleq\left[\cos \varphi_{ij}~ \sin\varphi_{ij}\right]^T.
\end{align}
In \eqref{eq:lambda_i}, $\beta_{{j}}$ denotes the effective bandwidth, and is given by
$\beta_{{j}}^2={\int_{-\infty}^{\infty}f^2|S_{{j}}(f)|^2df}\big{/}{\int_{-\infty}^{\infty}|S_{{j}}(f)|^2\,df}$, 
with $S_{{j}}(f)$ representing the Fourier transform of $s_{{j}}(t)$, and the path-overlap coefficient $\xi_{ij}$ is a non-negative number between zero and one, that is, $0\leq\xi_{ij}\leq1$ \cite{Moe_robust_power_allocation_2013}. In addition, $\varphi_{ij}$ in \eqref{eq:phiVec} denotes the angle between target node $i$ and anchor node $j$.

From \eqref{eq:CRLBineq} and \eqref{eq:CRLB1}, the CRLB for target node $i$ can be expressed as follows:
\begin{gather}\label{eq:CRLBi}
\CRLB_i={\rm{tr}}\left\{\bs{J}_i(\bs{x}_i,P_J)^{-1}\right\}
\end{gather}
where $\bs{J}_i(\bs{x}_i,P_J)$ is as in \eqref{eq:fisher}.


{\textit{\textbf{Remark 2:}} Even though the jammer noise received at different target nodes can be correlated in some cases, this does not have any effects on the formulation of the CRLB for each target node since the CRLB for a target node depends only on the signals received by that target node (cf.~\eqref{eq:fisher} and \eqref{eq:CRLBi}). In other words, since each target node is performing estimation of its own location, the jamming signals that affect the signals received by other target nodes are irrelevant for that target node.}

\section{Optimal Jammer Placement}\label{sec:OptJamPlace}

\subsection{Generic Formulation and Analysis}\label{sec:genericForm}

The aim is to determine the optimal location for the jammer node in order to increase the CRLBs of all the target nodes as much as possible. The CRLB is considered as a performance metric since it bounds the localization performance of a target node in terms of the mean-squared error \cite{Win_2014_IEEEnetw_PowerOpt,Leus_ICASSP10,Poor}. In particular, the minimum of the CRLBs of the target nodes is considered as the objective function to guarantee that all the target nodes have localization accuracy bounded by a certain limit. The proposed problem formulation is expressed, based on \eqref{eq:CRLBi}, as follows:
\begin{gather}\label{eq:OptProblem1}
\begin{split}
\underset{\bs{z}}{\mathrm{maximize}}~&\underset{i\in\{1,\ldots,N_T\}}\min~
{{\rm{tr}}\left\{\bs{J}_i(\bs{x}_i,P_J)^{-1}\right\}}
\\
{\mathrm{subject~to}}~&\|\bs{z}-\bs{x}_i\|\geq\varepsilon~,\quad i=1,\ldots,N_T
\end{split}
\end{gather}
where $\varepsilon>0$ denotes the lower limit for the distance between a target node and the jammer node, which is incorporated into the formulation since it may not be possible for the jammer node to get very close to target nodes in practical jamming scenarios (e.g., the jammer node may need to hide) \cite{a2}. 

Similarly to \cite{Win_2014_IEEEnetw_PowerOpt} and \cite{Molisch_2014_ICC_JointBWpowerAlloc}, the channel power gain between the jammer node and the $i$th target node is modeled as
\begin{gather}\label{eq:pathLoss}
|\gamma_{i{j}}|^2=\tilde{K}_i\left(\frac{d_0}{\|\bs{z}-\bs{x}_i\|}\right)^\nu~,
\end{gather}
for $\|\bs{z}-\bs{x}_i\|>d_0$, where $d_0$ is the reference distance for the antenna far-field, $\nu$ is the path-loss exponent (commonly between 2 and 4), and $\tilde{K}_i$ is a unitless constant that depends on antenna characteristics and average channel attenuation \cite{Goldsmith}. It is assumed that $\tilde{K}_i$'s, $d_0$, $\nu$, and $\varepsilon$ are known, and that $\varepsilon>d_0$.
(Also, the channel power gain between the jammer node and the $i$th target node is assumed to be constant during the reception of the signals from the anchor nodes.)
From \eqref{eq:pathLoss}, {the CRLB in \eqref{eq:CRLBi} can be stated, based on \eqref{eq:fisher}, as follows:}
\begin{gather}\label{eq:CRLBi_2}
{\CRLB_i
={\rm{tr}}\left\{\bs{J}_i(\bs{x}_i,P_J)^{-1}\right\}
=R_i\left(\frac{K_iP_J}{\|\bs{z}-\bs{x}_i\|^\nu}+\frac{N_0}{2}\right)}
\end{gather}
where $K_i\triangleq \tilde{K}_i(d_0)^{\nu}$ and
\begin{gather}\label{eq:ri}
R_i\triangleq {\rm{tr}} \left\{{\left[\sum_{j\in\mathcal{A}_i^L}\lambda_{ij}\bs{\phi}_{ij}\bs{\phi}^T_{ij}\right]^{-1}}\right\}.
\end{gather}
{Then,} the optimization problem in \eqref{eq:OptProblem1} can be expressed, {via \eqref{eq:CRLBi_2}}, as follows:\footnote{The jammer node is assumed to know the localization related parameters so that it can solve the optimization problem in \eqref{eq:OptProblem2}. Although this information may not completely be available to the jammer node in practical scenarios, this assumption is made for two purposes: (i) to obtain initial results which can form a basis for further studies on the problem of optimal jammer placement in wireless localization systems, (ii) to derive theoretical limits on the best achievable performance of the jammer node (if the jammer node is smart and can learn all the related parameters, the localization accuracy provided in this study is achieved; otherwise, the localization accuracy is bounded by the provided results).}
\begin{gather}\label{eq:OptProblem2}
\begin{split}
\underset{\bs{z}}{\mathrm{maximize}}~&\underset{i\in\{1,\ldots,N_T\}}\min~
R_i\left(\frac{K_iP_J}{\|\bs{z}-\bs{x}_i\|^\nu}+\frac{N_0}{2}\right)\\
{\mathrm{subject~to}}~&\|\bs{z}-\bs{x}_i\|\geq\varepsilon~,\quad i=1,\ldots,N_T
\end{split}
\end{gather}
{Since the jammer node is assumed to know the localization related parameters in this formulation, a performance benchmark is provided for the jamming of wireless localization systems, which corresponds to the best achievable performance for the jammer node and the worst-case scenario for the wireless localization network. Hence, based on the results in this study, a wireless localization network can specify the maximum amount of performance degradation that can be caused by a jammer node and take certain precautions accordingly (see Section~\ref{sec:Conc}).}

The problem in \eqref{eq:OptProblem2} is non-convex; hence, convex optimization tools cannot be employed to obtain the optimal location of the jammer node. Therefore, an exhaustive search over the feasible locations for the jammer node may be required in general.
However, some theoretical results are obtained in the following in order to simplify the optimization problem in \eqref{eq:OptProblem2} under various conditions.

\textit{\textbf{Proposition 1:} If there exists a target node, say the $\ell$th one, that satisfies the following inequality,
\begin{gather}\label{eq:Prop1_ineq}
R_\ell\bigg(\frac{K_\ell P_J}{\varepsilon^\nu}+\frac{N_0}{2}\bigg)
\leq\underset{i\ne\ell}{\underset{i\in\{1,\ldots,N_T\}}
\min}
R_i\bigg(\frac{K_iP_J}{(\|\bs{x}_i-\bs{x}_\ell\|+\varepsilon)^\nu}+\frac{N_0}{2}\bigg)
\end{gather}
and if set $\{\bs{z}\,:\,\|\bs{z}-\bs{x}_\ell\|=\varepsilon~\&~\|\bs{z}-\bs{x}_i\|\geq\varepsilon,\, i=1,\ldots,\ell-1,\ell+1,\ldots,N_T\}$ is non-empty, then the solution of \eqref{eq:OptProblem2}, denoted by $\bs{z}^{\rm{opt}}$, satisfies $\|\bs{z}^{\rm{opt}}-\bs{x}_\ell\|=\varepsilon$; that is, the jammer node is placed at a distance of $\varepsilon$ from the $\ell$th target node.}
\begin{proof} {See Appendix~\ref{app:Prop1}.} \end{proof}

Proposition~1 presents a scenario in which the jammer node must be as close to a certain target node (denoted by target node $\ell$ in the proposition) as possible in order to maximize the minimum of the CRLBs of the target nodes. In this scenario, the feasible set for the jammer location is significantly reduced, which simplifies the search space for the optimization problem in \eqref{eq:OptProblem2}.


In order to specify another scenario in which the solution of \eqref{eq:OptProblem2} can be obtained in a simplified manner, consider the optimization problem in \eqref{eq:OptProblem2} in the presence of two target nodes $\ell_1$ and $\ell_2$ only; that is,
\begin{gather}\label{eq:OptProblem_ij}
\begin{split}
\underset{\bs{z}}{\mathrm{maximize}}~&\underset{i\in\{\ell_1,\,\ell_2\}}\min~
R_i\left(\frac{K_iP_J}{\|\bs{z}-\bs{x}_i\|^\nu}+\frac{N_0}{2}\right)\\
{\mathrm{subject~to}}~&\|\bs{z}-\bs{x}_{\ell_1}\|\geq\varepsilon\,,~\|\bs{z}-\bs{x}_{\ell_2}\|\geq\varepsilon
\end{split}
\end{gather}
where $\ell_1,\ell_2\in\{1,\ldots,N_T\}$ and $\ell_1\ne\ell_2$. Let $\bs{z}_{\ell_1,\ell_2}^{\rm{opt}}$ and $\CRLB_{\ell_1,\ell_2}$ denote the optimizer and the optimal value of \eqref{eq:OptProblem_ij}, respectively. (In the next section, the solution in the presence of two target nodes is investigated in detail.)
Then, the following proposition characterizes the solution of \eqref{eq:OptProblem2} under certain conditions.

\textit{\textbf{Proposition 2:} Let $\CRLB_{k,i}$ be the minimum of $\CRLB_{\ell_1,\ell_2}$ for $\ell_1,\ell_2\in\{1,\ldots,N_T\}$ and $\ell_1\ne\ell_2$, and let $\bs{z}_{k,i}^{\rm{opt}}$ denote
the corresponding jammer location (i.e., the optimizer of \eqref{eq:OptProblem_ij} for $\ell_1=k$ and $\ell_2=i$). Then, an optimal jammer location obtained from \eqref{eq:OptProblem2} is equal to $\bs{z}_{k,i}^{\rm{opt}}$ if $\bs{z}_{k,i}^{\rm{opt}}$ is an element of set $\big\{\bs{z}\,:\,\|\bs{z}-\bs{x}_m\|\geq\varepsilon,\, m\in\{1,\ldots,N_T\}\setminus\{k,i\}\big\}$ and}
\begin{gather}\label{eq:prop2}
R_m\left(\frac{K_mP_J}{\|\bs{z}_{k,i}^{\rm{opt}}-\bs{x}_m\|^\nu}+\frac{N_0}{2}\right)\geq\CRLB_{k,i}
\end{gather}
\textit{for} $m\in\{1,\ldots,N_T\}\setminus\{k,i\}$.
\begin{proof}
From \eqref{eq:OptProblem2} and \eqref{eq:OptProblem_ij}, it is noted that $\CRLB_{k,i}$, defined in the proposition, provides an upper bound for the problem in \eqref{eq:OptProblem2}. If the conditions in \eqref{eq:prop2} are satisfied, the objective function in \eqref{eq:OptProblem2} becomes equal to the upper bound, $\CRLB_{k,i}$, for $\bs{z}=\bs{z}_{k,i}^{\rm{opt}}$. Therefore, if $\bs{z}_{k,i}^{\rm{opt}}$ satisfies the distance constraints (i.e., if it is feasible for \eqref{eq:OptProblem2}), it becomes the  solution of \eqref{eq:OptProblem2}.
\end{proof}

Proposition~2 specifies a scenario in which the optimal jammer location is mainly determined by two of the target nodes since the others have larger CRLBs when the jammer node is placed at the optimal location according to those two jammer nodes only. In such a scenario, the optimal jammer location can be found easily, as the solution of \eqref{eq:OptProblem_ij} is simple to obtain (in comparison to \eqref{eq:OptProblem2}), which is investigated in the following section.

\subsection{Special Case: Two Target Nodes}\label{sec:2target}

In the case of two target nodes, the solution of \eqref{eq:OptProblem2} can easily be obtained based on the following 
result.

\textit{\textbf{Proposition 3:} For the case of two target nodes (i.e., $N_T=2$), the solution $\bs{z}^{\rm{opt}}$ of \eqref{eq:OptProblem2} satisfies one of the following conditions:}

\textit{\textbf{(i)} if $\|\bs{x}_1-\bs{x}_2\| < 2\,\varepsilon$, then} $\|\bs{z}^{\rm{opt}}-\bs{x}_1\|=\|\bs{z}^{\rm{opt}}-\bs{x}_2\|=\varepsilon$.

\textit{\textbf{(ii)} otherwise,}

\indent\indent\textit{\textbf{(a)} if $R_1\left(\frac{K_1P_J}{\varepsilon^\nu}+\frac{N_0}{2}\right)\leq R_2\left(\frac{K_2P_J}{(\|\bs{x}_1-\bs{x}_2\|-\varepsilon)^\nu}+\frac{N_0}{2}\right)$,
then
$\|\bs{z}^{\rm{opt}}-\bs{x}_1\|=\varepsilon$
and}
$\|\bs{z}^{\rm{opt}}-\bs{x}_2\|=\|\bs{x}_1-\bs{x}_2\|-\varepsilon$.

\indent\indent\textit{\textbf{(b)} if $R_2\left(\frac{K_2P_J}{\varepsilon^\nu}+\frac{N_0}{2}\right)\leq R_1\left(\frac{K_1P_J}{(\|\bs{x}_1-\bs{x}_2\|-\varepsilon)^\nu}+\frac{N_0}{2}\right)$,
then
$\|\bs{z}^{\rm{opt}}-\bs{x}_1\|=\|\bs{x}_1-\bs{x}_2\|-\varepsilon$
and}
$\|\bs{z}^{\rm{opt}}-\bs{x}_2\|=\varepsilon$.

\indent\indent\textit{\textbf{(c)} otherwise,
$\|\bs{z}^{\rm{opt}}-\bs{x}_1\|=d^*$
and
$\|\bs{z}^{\rm{opt}}-\bs{x}_2\|=\|\bs{x}_1-\bs{x}_2\|-d^*$,
where $d^*$ is the unique solution of the following equation over $d\in(\varepsilon,\,\|\bs{x}_1-\bs{x}_2\|-\varepsilon)$.}
\begin{gather}\label{eq:equalizer2}
R_1\left(\frac{K_1P_J}{d^{\,\nu}}+\frac{N_0}{2}\right)
=R_2\left(\frac{K_2P_J}{(\|\bs{x}_1-\bs{x}_2\|-d)^\nu}+\frac{N_0}{2}\right)
\end{gather}
\begin{proof} {See Appendix~\ref{app:Prop3}.} \end{proof}

Based on Proposition~3, the optimal location of the jammer node can be specified for $N_T=2$ as follows: If the distance between the target nodes is smaller than $2\,\varepsilon$, then the jammer node is located at one of the two intersections of the circles around the target nodes with radius of $\varepsilon$ each.
Otherwise, the jammer node is always on the straight line that connects the two target nodes; that is, $\|\bs{z}^{\rm{opt}}-\bs{x}_1\|+\|\bs{z}^{\rm{opt}}-\bs{x}_2\|=\|\bs{x}_2-\bs{x}_1\|$. In this case, depending on the CRLB values, the jammer node can be either at a distance of $\varepsilon$ from one of the target nodes (the one with the lower CRLB) or at larger distances than $\varepsilon$ from both of the target nodes. In the first scenario, the optimal jammer position is simply obtained as $\bs{z}^{\rm{opt}}=\bs{x}_i+(\bs{x}_k-\bs{x}_i)\varepsilon/\|\bs{x}_k-\bs{x}_i\|$ when the jammer node is at a distance of $\varepsilon$ from the $i$th target node. In the second scenario, an \textit{equalizer} solution is observed as the CRLBs are equated, and the optimal jammer location is calculated as $\bs{z}^{\rm{opt}}=\bs{x}_1+(\bs{x}_2-\bs{x}_1)d^*/\|\bs{x}_2-\bs{x}_1\|$, where
$d^*$ is obtained from \eqref{eq:equalizer2}.

\subsection{Special Case: Infinitesimally Small $\varepsilon$}\label{sec:3target}

In this section, the optimal location of the jammer node is investigated for $N_T\geq3$ in the absence of constraints on the distances between the jammer node and the target nodes; that is, it is assumed that the constraints in \eqref{eq:OptProblem2} are ineffective. In this scenario, various theoretical results can be obtained related to the optimal location for the jammer node.

\textit{\textbf{Remark {3}:} The ineffectiveness of the distance constraints can naturally arise in some cases due to the max-min nature of the problem; that is, the solution of the problem in \eqref{eq:OptProblem2} can be the same in the presence and absence of the constraints (see Section~\ref{sec:Simu} for examples). In addition, for applications in which small (e.g., `nano size'  \cite{a12}) jammer nodes with low powers are employed, the jammer node becomes difficult to detect; hence, it can be placed closely to the target nodes, leading to a low value of $\varepsilon$ in \eqref{eq:OptProblem2}.}


First, the following 
result is obtained to restrict the possible region for the optimal jammer location.

\textit{\textbf{Proposition 4:} Suppose that $N_T\geq3$ and $\varepsilon\rightarrow0$. Then, the optimal location of the jammer node lies on the convex hull formed by the locations of the target nodes.}
\begin{proof}
Let ${\mathcal{H}}$ denote the convex hull formed by the locations of the target nodes; that is, ${\mathcal{H}}={\rm{Conv}}(\bs{x}_1,\ldots,\bs{x}_{N_T})=\big\{\sum_{i=1}^{N_T}\upsilon_i\,\bs{x}_i\,|\,\sum_{i=1}^{N_T}\upsilon_i=1,\,\upsilon_i\geq0,\,i=1,\ldots,N_T\big\}$. By definition, ${\mathcal{H}}$ is a nonempty closed convex set. Let $\bs{z}_1$ be any point outside ${\mathcal{H}}$. Then, by the projection theorem \cite{Bertsekas2}, there exits a unique vector $\bs{z}_2$ in ${\mathcal{H}}$ that is closest to $\bs{z}_1$; that is, $\bs{z}_2=\rm{argmin}_{\bs{z}\in{\mathcal{H}}}\,\|\bs{z}-\bs{z}_1\|$ (i.e., $\bs{z}_2$ is the \textit{projection} of $\bs{z}_1$ onto ${\mathcal{H}}$). The projection theorem also states that $\bs{z}_2$ is the projection of $\bs{z}_1$ onto ${\mathcal{H}}$ if and only if $(\bs{z}_1-\bs{z}_2)^T(\bs{z}_3-\bs{z}_2)\leq0$ for all $\bs{z}_3\in{\mathcal{H}}$ \cite{Bertsekas2}. This condition can also be stated as
\begin{gather}\label{eq:Lem1_1}
\bs{z}_1^T\bs{z}_3-\bs{z}_1^T\bs{z}_2-\bs{z}_2^T\bs{z}_3+\|\bs{z}_2\|^2\leq0\,.
\end{gather}
Multiplying the terms in \eqref{eq:Lem1_1} by 2 and moving some of the terms to the other side, the following inequality is obtained:
\begin{gather}\label{eq:Lem1_2}
2\bs{z}_1^T\bs{z}_2-\|\bs{z}_2\|^2\geq2\bs{z}_1^T\bs{z}_3+\|\bs{z}_2\|^2-2\bs{z}_2^T\bs{z}_3~.
\end{gather}
Since $\bs{z}_1\notin{\mathcal{H}}$ and $\bs{z}_2\in{\mathcal{H}}$, $\|\bs{z}_1-\bs{z}_2\|>0$ is satisfied, which is equivalent to $\|\bs{z}_1\|^2>2\bs{z}_1^T\bs{z}_2-\|\bs{z}_2\|^2$. Then, from \eqref{eq:Lem1_2}, the following relation is derived:
\begin{gather}\label{eq:Lem1_3}
\|\bs{z}_1\|^2>2\bs{z}_1^T\bs{z}_3+\|\bs{z}_2\|^2-2\bs{z}_2^T\bs{z}_3~.
\end{gather}
Adding $\|\bs{z}_3\|^2$ to both sides of the inequality in \eqref{eq:Lem1_3}, and rearranging the terms, the following distance relation is achieved:
\begin{gather}\label{eq:Lem1_4}
\|\bs{z}_1-\bs{z}_3\|>\|\bs{z}_2-\bs{z}_3\|
\end{gather}
for all $\bs{z}_3\in{\mathcal{H}}$. Hence, for any point $\bs{z}_1$ outside ${\mathcal{H}}$, its projection onto ${\mathcal{H}}$, denoted by $\bs{z}_2$, is closer to any point $\bs{z}_3$ on ${\mathcal{H}}$. Therefore, the optimal jammer location cannot be outside the convex hull ${\mathcal{H}}$ formed by the locations of the target nodes as the CRLB for each target node is inversely proportional to the distance between the jammer and the target nodes.
\end{proof}

\begin{figure}
\vspace{-0.1cm}
\center
  \includegraphics[width=60mm]{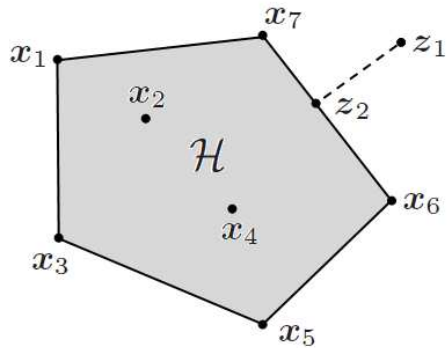}
\vspace{-0.2cm}
\caption{A scenario with $N_T=7$ target nodes, where ${\mathcal{H}}$ denotes the convex hull formed by the locations of the target nodes (the gray area). Point $\bs{z}_2$ is the projection of $\bs{z}_1$ onto ${\mathcal{H}}$.}
 \label{fig:prop4}
\vspace{-0.3cm}
\end{figure}


The statement in Proposition~4 is illustrated in Fig.~\ref{fig:prop4}. As stated in the proof of the proposition, for each location $\bs{z}_1$ outside the convex hull ${\mathcal{H}}$ (formed by the locations of the target nodes), its projection $\bs{z}_2$ onto ${\mathcal{H}}$ is closer to all the locations on ${\mathcal{H}}$, hence, to all the target nodes. Therefore, the optimal jammer location must be always on the convex hull generated by the target nodes.

{In \cite{ConHullPos}, a semidefinite programming (SDP) relaxation based method is proposed for localization of target nodes in the absence of jamming, and it is observed that target nodes should be in the convex hull of the anchor nodes in order to perform accurate localization. However, this observation is different from the result in Proposition~4 in terms of both the considered problem and the employed proof technique.}

Towards the aim of characterizing the optimal jammer location for $N_T>3$, the scenario with $N_T=3$ is investigated first. Consider a network with target nodes $\ell_1$, $\ell_2$, and $\ell_3$ (i.e., $N_T=3$). The max-min CRLB in the absence of distance constraints is defined as
\begin{gather}\label{eq:optProb_3tar_NoConstr}
\CRLB_{\ell_1,\ell_2,\ell_3}\triangleq\max_{\bs{z}}\,\min_{{m\in\{\ell_1,\ell_2,\ell_3\}}}\,\CRLB_m(\bs{z})
\end{gather}
where $\CRLB_m(\bs{z})$ is given by (cf.~\eqref{eq:OptProblem2})
\begin{gather}\label{eq:CRLBm}
\CRLB_m(\bs{z})\triangleq R_m\left(\frac{K_mP_J}{\|\bs{z}-\bs{x}_m\|^\nu}+\frac{N_0}{2}\right)\,. \end{gather}
According to Proposition~4, the optimal jammer location lies on the triangle formed by the locations of target nodes $\ell_1$, $\ell_2$, and $\ell_3$. In particular, the jammer node can be either inside the triangle or on the boundary of the triangle.\footnote{If the target nodes are co-linear, then the jammer node resides on the boundary of the `triangle', which in fact reduces to a straight line segment.} For the former case, the following proposition presents the equalizer nature of the optimal solution.

\textit{\textbf{Proposition 5:} Consider a network with three target nodes (i.e., $N_T=3$). If the optimal jammer location obtained from \eqref{eq:optProb_3tar_NoConstr} belongs to the interior of the convex hull (triangle) formed by the locations of the target nodes, then the CRLBs for the target nodes are equalized by the optimal solution.}
\begin{proof} {See Appendix~\ref{app:Prop5}.} \end{proof}

Based on Proposition~5, it is concluded that if the optimal jammer location obtained from \eqref{eq:optProb_3tar_NoConstr} belongs to the interior of the convex hull (triangle) formed by the three target nodes, then the resulting CRLBs for the target nodes are all equal.
To investigate the scenario in which the optimal jammer location is on the boundary of the triangle formed by target nodes $\ell_1$, $\ell_2$, and $\ell_3$, $\CRLB_{m,n}$ is defined as
\begin{gather}\label{eq:optProb_2tar_NoConstr}
\CRLB_{m,n}\triangleq\max_{\bs{z}}\,\min\{\CRLB_m(\bs{z}),\CRLB_n(\bs{z})\}
\end{gather}
where $\CRLB_m(\bs{z})$ and $\CRLB_n(\bs{z})$ are given by \eqref{eq:CRLBm}. First, based on Proposition~3, the following result is obtained for two target nodes ($N_T=2$) in the absence of distance constraints (i.e., $\varepsilon\rightarrow0$).

\textit{\textbf{Corollary 1:} For two target nodes and without distance constraints on the location of the jammer node, the optimal jammer location (see \eqref{eq:optProb_2tar_NoConstr}) is on the straight line segment that connects the target nodes, and the CRLBs for the target nodes are equalized by the optimal solution.}
\begin{proof}
Consider Proposition~3 with $\varepsilon\rightarrow0$. Then, the only possible scenario is $(ii)$--$(c)$, which results in an equalizer solution with the jammer node being located on the straight line segment that connects the target nodes.
\end{proof}

Then, the following proposition characterizes the scenario in which the optimal jammer location according to \eqref{eq:optProb_3tar_NoConstr} is on the boundary of the triangle formed by the target nodes.

\textit{\textbf{Proposition 6:} Consider a network with target nodes $\ell_1$, $\ell_2$, and $\ell_3$, and suppose that $\CRLB_{\ell_1,\ell_2}$ is the minimum of $\{\CRLB_{\ell_1,\ell_2},\CRLB_{\ell_1,\ell_3},\CRLB_{\ell_2,\ell_3}\}$
(see \eqref{eq:optProb_2tar_NoConstr}).\footnote{It is possible to extend the results to scenarios in which $\CRLB_{\ell_1,\ell_2}$ is not a unique minimum.} Also, let $\bs{z}_{\ell_1,\ell_2}^{\rm{opt}}$ represent the optimizer of \eqref{eq:optProb_2tar_NoConstr} for $m=\ell_1$ and $n=\ell_2$. Then, the optimal jammer location obtained from \eqref{eq:optProb_3tar_NoConstr} satisfies the following properties:}

\textit{\textbf{a)} If the optimal jammer location is on the boundary of the triangle formed by target nodes $\ell_1$, $\ell_2$, and $\ell_3$, then the optimizer of \eqref{eq:optProb_3tar_NoConstr} is equal to $\bs{z}_{\ell_1,\ell_2}^{\rm{opt}}$, and the CRLBs for target nodes $\ell_1$ and $\ell_2$ are equalized by the optimal solution; that is, $\CRLB_{\ell_1}(\bs{z}_{\ell_1,\ell_2}^{\rm{opt}})=\CRLB_{\ell_2}(\bs{z}_{\ell_1,\ell_2}^{\rm{opt}})$.}

\textit{\textbf{b)} The optimal location for the jammer node is on the boundary of the convex hull (triangle) formed by target nodes $\ell_1$, $\ell_2$, and $\ell_3$ if and only if}
\begin{gather}\label{eq:Prop6b}
\|\bs{x}_{\ell_3}-\bs{z}_{\ell_1,\ell_2}^{\rm{opt}}\|\leq
\sqrt[\nu]{P_JK_{\ell_3}\left(\frac{\CRLB_{\ell_1,\ell_2}}{R_{\ell_3}}-\frac{N_0}{2}\right)^{-1}}\,.
\end{gather}

\begin{proof} {See Appendix~\ref{app:Prop6}.} \end{proof}

Proposition~6 presents a necessary and sufficient condition for the optimal jammer location to be on the boundary of the convex hull (triangle) formed by the three target nodes (see \eqref{eq:Prop6b}) in the absence of distance constraints. To utilize the results in Proposition~6, $\CRLB_{\ell_1,\ell_2}$, $\CRLB_{\ell_1,\ell_3}$, and $\CRLB_{\ell_2,\ell_3}$ are calculated from \eqref{eq:optProb_2tar_NoConstr}, and the condition in \eqref{eq:Prop6b} is checked. If the condition holds, the optimal location for the jammer node is obtained as specified in Part a) of the proposition, which results in equalization of the CRLBs for (at least) two of the target nodes. Otherwise, the optimal location for the jammer node belongs to the interior of the convex hull, and the result in Proposition~5 applies.

Based on Propositions 4--6, the following result is obtained to characterize the optimal location for the jammer node for $N_T>3$ and in the absence of distance constraints.

\textit{\textbf{Proposition 7:} Suppose that $N_T>3$ and $\varepsilon\rightarrow0$. Let the max-min CRLB in the presence of target nodes $\ell_1$, $\ell_2$, and $\ell_3$ only be denoted by $\CRLB_{\ell_1,\ell_2,\ell_3}$, which is as expressed in \eqref{eq:optProb_3tar_NoConstr}.
Assume that target nodes $i$, $j$, and $k$ achieve the minimum of $\CRLB_{\ell_1,\ell_2,\ell_3}$ for $\ell_1,\ell_2,\ell_3\in\{1,\ldots,N_T\}$ and $\ell_1\ne\ell_2\ne\ell_3$, and let $\bs{z}^{\rm{opt}}_{i,j,k}$ denote the optimizer of \eqref{eq:optProb_3tar_NoConstr} corresponding to $\CRLB_{i,j,k}$; that is, for $(\ell_1,\ell_2,\ell_3)=(i,j,k)$.
Then, the optimal location for the jammer node (i.e., the optimizer of \eqref{eq:OptProblem2} in the absence of the distance constraints) is equal to $\bs{z}^{\rm{opt}}_{i,j,k}$, and at least two of the CRLBs of the target nodes are equalized by the optimal solution.}
\begin{proof} {See Appendix~\ref{app:Prop7}.} \end{proof}

The significance of Proposition~7 is related to the statement that the optimal location of the jammer node is determined by no more than three of the target nodes for infinitesimally small $\varepsilon$. In addition, when the optimal location of the jammer node is obtained based on Proposition~7 as $\bs{z}^{\rm{opt}}_{i,j,k}$, it also becomes the solution of \eqref{eq:OptProblem2} if $\bs{z}^{\rm{opt}}_{i,j,k}$ is an element of $\{\bs{z}\,|\,\|\bs{z}-\bs{x}_i\|\geq\varepsilon\,,~i=1,\ldots,N_T\}$. Otherwise, \eqref{eq:OptProblem2} results in a different solution.




Finally, the following corollary is obtained based on Propositions 5--7.

\textit{\textbf{Corollary 2:} Consider the scenario in Proposition~7 and suppose that the optimal location for the jammer node, $\bs{z}^{\rm{opt}}_{i,j,k}$, belongs to the interior of the convex hull formed by target nodes $i$, $j$, and $k$. In addition, let $\CRLB_{i,j}$ be the minimum of $\CRLB_{i,j}$, $\CRLB_{i,k}$, and $\CRLB_{j,k}$, which are as defined in \eqref{eq:optProb_2tar_NoConstr}, and let $\bs{z}_{i,j}^{\rm{opt}}$ represent the jammer location corresponding to $\CRLB_{i,j}$. Then, $\bs{z}^{\rm{opt}}_{i,j,k}$ cannot be inside any of the circles centered at target nodes $i$, $j$, and $k$ with radii $\|\bs{x}_i-\bs{z}_{i,j}^{\rm{opt}}\|$, $\|\bs{x}_j-\bs{z}_{i,j}^{\rm{opt}}\|$, and $d_{\rm{thr}}$, respectively, where}
\begin{gather}\label{eq:Coro2}
d_{\rm{thr}}\triangleq\sqrt[\nu]{P_JK_k\left(\frac{\CRLB_{i,j}}{R_k}-\frac{N_0}{2}\right)^{-1}}\,.
\end{gather}

\begin{figure}
\vspace{-0.1cm}
\center
  \includegraphics[width=70mm]{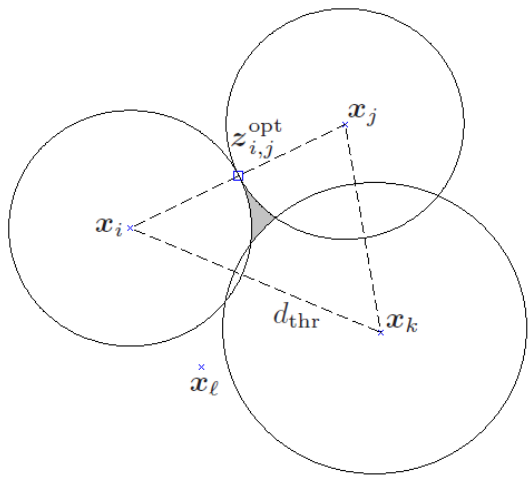}
  \vspace{-0.2cm}
\caption{The scenario in Corollary~2, where the optimal location for the jammer node corresponds to a point in the shaded (gray) area.}
 \label{fig:cor2}
 \vspace{-0.2cm}
\end{figure}

The statement in Corollary~2 is illustrated in Fig.~\ref{fig:cor2}. According to Corollary~2, the jammer node cannot be inside any of the three circles shown in the figure, and the only feasible region is the shaded area. This corollary is useful to reduce the search region for the optimal location of the jammer node.


{Based on the theoretical results in this section, the following algorithm can be proposed for calculating the optimal location of the jammer node, $\bs{z}^{\rm{opt}}$, for the generic problem in \eqref{eq:OptProblem2}:
\begin{itemize}
\item If $N_T=1$, $\bs{z}^{\rm{opt}}$ can be chosen to be any point at a distance of $\varepsilon$ from the target node.
\item If $N_T=2$, $\bs{z}^{\rm{opt}}$ can be obtained from Proposition~3, which presents either a closed-form solution, or a solution based on a simple one-dimensional search (see \eqref{eq:equalizer2}).
\item If $N_T\geq3$,
\begin{itemize}
\item If the conditions in Proposition~1 hold, $\bs{z}^{\rm{opt}}$ is at a distance of $\varepsilon$ from a specific target node.
\item If the conditions in Proposition~2 hold, $\bs{z}^{\rm{opt}}$ is determined by two of the target nodes, as described in Proposition~3.
\item Otherwise,
    \begin{itemize}
    \item For each distinct group of three target nodes, say $\ell_1$, $\ell_2$, and $\ell_3$,
        \begin{itemize}
        \item Calculate the pairwise CRLBs in \eqref{eq:optProb_2tar_NoConstr} considering the equalizer property in Corollary~1, and determine the minimum of them, say $\CRLB_{\ell_1,\ell_2}$.
        \item If the condition in \eqref{eq:Prop6b} of Proposition~6 holds, set $\CRLB_{\ell_1,\ell_2,\ell_3}$ to $\CRLB_{\ell_1,\ell_2}$.
        \item Otherwise, obtain $\CRLB_{\ell_1,\ell_2,\ell_3}$ from \eqref{eq:optProb_3tar_NoConstr} under the equalizer constraint specified in Proposition~5.
        \end{itemize}
\item Determine the minimum of the $\CRLB_{\ell_1,\ell_2,\ell_3}$ terms and the corresponding optimal location, $\bs{z}^{\rm{opt}}_{\rm{unc}}$ (i.e., the optimal location in the absence of distance constraints).
\item If $\bs{z}^{\rm{opt}}_{\rm{unc}}$ is feasible according to \eqref{eq:OptProblem2}, then $\bs{z}^{\rm{opt}}=\bs{z}^{\rm{opt}}_{\rm{unc}}$. Otherwise, solve \eqref{eq:OptProblem2} directly to obtain $\bs{z}^{\rm{opt}}$.
\end{itemize}
\end{itemize}
\end{itemize}
It should be noted that the solution of \eqref{eq:OptProblem2} requires a two-dimensional search over the set of feasible locations for the jammer node. On the other hand, the algorithm based on Propositions~5--7 involves $\binom{N_T}{3}$ optimization problems, each of which is over a one-dimensional space due to the equalizer properties in the propositions.} {In the worst case where \eqref{eq:OptProblem2} is solved exhaustively, $N_FN_T$ evaluations of the CRLB expression in \eqref{eq:CRLBi_2} is required, with $N_F$ denoting the number of feasible locations in the environment (considering a certain resolution for the search). On the other hand, in the best case, Proposition~1 can be applied and the optimal jammer location can be obtained with no more than $(N_T)^2$ CRLB evaluations (see \eqref{eq:Prop1_ineq}).}

\section{{Extensions}}\label{sec:Exten}

{In practical localization systems, an anchor node can be connected to a target node if the signal-to-noise ratio (SNR) at the receiver of the target node is larger than a certain threshold. Since the jammer node degrades the SNRs at the target nodes, it may be possible in some cases that the set of anchor nodes that are connected to a target node can change with respect to the location of the jammer node. In order to incorporate such cases, the problem formulation in the previous sections can be generalized as follows: Let $\mathcal{A}_i$ in Section~\ref{sec:SysModel} now represent the set of anchor nodes that are connected to the $i$th target node \textit{in the absence of jamming}. In addition, let $\SNR_{ij}$ denote the SNR of the received signal coming to target node~$i$ from anchor node~$j$, which can be expressed as $\SNR_{ij}=E_{ij}/({K_iP_J}/{\|\bs{z}-\bs{x}_i\|^\nu}+{N_0}/{2})$, where $E_{ij}$ is the energy of the signal coming from anchor node~$j$ (i.e., the energy of the first term in \eqref{eq:model}) and ${K_iP_J}/{\|\bs{z}-\bs{x}_i\|^\nu}+{N_0}/{2}$ is the sum of the spectral density levels of the jammer noise (cf.~\eqref{eq:pathLoss}) and the measurement noise. Then, the condition that $\SNR_{ij}$ is above a threshold, $\SNRt$, can be expressed, after some manipulation, as follows:
\begin{gather}\label{eq:SNRcond}
\|\bs{z}-\bs{x}_i\|>\left(\frac{K_iP_J}{E_{ij}/\SNRt-N_0/2}\right)^{1/\nu}\triangleq d_{ij}^{\rm{lim}}
\end{gather}
for $i\in\{1,\ldots,N_T\}$ and $j\in\mathcal{A}_i$, where $E_{ij}/\SNRt>N_0/2$ holds for $j\in\mathcal{A}_i$ by definition. The inequality in \eqref{eq:SNRcond} states that if the distance between the jammer node and target node $i$ is larger than a critical distance $d_{ij}^{\rm{lim}}$, then target node~$i$ can utilize the signal coming from anchor node $j$; otherwise, target node~$i$ cannot communicate with anchor node~$j$. In this scenario, the CRLB expressions can be updated by incorporating these conditions into \eqref{eq:fisher} as follows:
\begin{align}\label{eq:fisherNew}
\bs{J}_i(\bs{x}_i,P_J)=\sum_{j\in\mathcal{A}_i^L}\frac{\lambda_{ij}
{\mathbb{I}}_{\{\|\bs{z}-\bs{x}_i\|>d_{ij}^{\rm{lim}}\}}}{N_0/2
+P_J|\gamma_{ij}|^2}\,\bs{\phi}_{ij}\bs{\phi}^T_{ij}
\end{align}
where ${\mathbb{I}}$ denotes an indicator function, which is equal to one when the condition is satisfied and zero otherwise. From \eqref{eq:fisherNew}, the CRLB in \eqref{eq:CRLBi_2} and \eqref{eq:ri} can be expressed, via \eqref{eq:pathLoss}, as
\begin{gather}\label{eq:CRLBi_2_New}
\CRLB_i(d_i)=R_i(d_i)\left({K_iP_J}/{(d_i)^\nu}+{N_0}/{2}\right)
\end{gather}
where $d_i\triangleq\|\bs{z}-\bs{x}_i\|$ and
\begin{gather}\label{eq:ri_New}
R_i(d_i)\triangleq {\rm{tr}} \left\{{\left[\sum_{j\in\mathcal{A}_i^L}\lambda_{ij}{\mathbb{I}}_{\{d_i>d_{ij}^{\rm{lim}}\}}\bs{\phi}_{ij}\bs{\phi}^T_{ij}\right]^{-1}}\right\}.
\end{gather}
Based on the new CRLB expression in \eqref{eq:CRLBi_2_New} and \eqref{eq:ri_New}, the extensions of the theoretical results in Section~\ref{sec:OptJamPlace} can be investigated as follows:
Proposition~1 can directly be applied by replacing the condition in \eqref{eq:Prop1_ineq} with the following:
\begin{gather}\label{eq:Prop1_ineq_New}
\CRLB_{\ell}(\varepsilon)\leq\underset{i\ne\ell}{\underset{i\in\{1,\ldots,N_T\}}
\min}\CRLB_i(\|\bs{x}_i-\bs{x}_\ell\|+\varepsilon).
\end{gather}
Similarly, Proposition~2 can be employed by using the following inequality instead of \eqref{eq:prop2}:
$\CRLB_m(\|\bs{z}_{k,i}^{\rm{opt}}-\bs{x}_m\|)\geq \CRLB_{k,i}$,
where $\CRLB_{k,i}$ denotes the solution of \eqref{eq:OptProblem_ij} when $R_i$ in the objective function is as defined in \eqref{eq:ri_New}. Regarding Proposition~3, Part~$(i)$ directly applies, and Part~$(ii)$--$(a)$ and Part~$(ii)$--$(b)$ are valid when the definition of $R_i$ is updated. However, Part~$(ii)$--$(c)$ does not directly apply since equalization may not be possible due to the discontinuous nature of the CRLB expression in \eqref{eq:CRLBi_2_New} and \eqref{eq:ri_New}. Hence, in this scenario, instead of \eqref{eq:equalizer2}, the following conditions should be employed for $d^*$:
\begin{gather}
\begin{split}
\CRLB_1(d)&\geq \CRLB_2(\|\bs{x}_1-\bs{x}_2\|-d)\,~\textrm{for }d<d^*\\
\CRLB_1(d)&\leq \CRLB_2(\|\bs{x}_1-\bs{x}_2\|-d)\,~\textrm{for }d>d^*
\end{split}
\end{gather}
Proposition~4 can also be directly applied under the assumption that the jammer node cannot disable all the target nodes from a location outside the convex hull (that is, the minimum CRLB of the target nodes should be finite for all jammer locations outside the convex hull). Regarding Propositions~5--7, the continuity property of the CRLB plays an important role for proving the results in these propositions. Therefore, they do not apply in general for the CRLB expression in \eqref{eq:CRLBi_2_New} and \eqref{eq:ri_New}. To extend the results in Propositions~5--7, a continuous approximation of the CRLB expression can be considered. From \eqref{eq:ri_New}, it is noted that the CRLB can have finitely many discontinuities, the number of which is determined by the number of anchor nodes. Hence, by approximating the CRLB from below (so that it is still a lower bound) around those discontinuities leads to an approximate formulation for which the results in Propositions~5--7 can be applied. Investigation of such approximations and their practical implications are considered as a direction for future work.}

{\textit{\textbf{Remark~4}}: The theoretical results in this manuscript are valid not only for the CRLB expressions that are derived based on the considered system model but also for any localization accuracy metric that satisfies the following properties: $(i)$ The localization accuracy improves as the distance between the jammer node and the target node increases. $(ii)$ The localization accuracy metric is a continuous function of the distance between the jammer node and the target node. In particular, Propositions 1, 2, 3, 4 and Corollary~1 can directly be extended when condition $(i)$ is satisfied. On the other hand, the results in Propositions 5, 6, 7 and Corollary~2 are valid when both condition $(i)$ and $(ii)$ are satisfied. Since the first property should hold for any reasonable average performance metric for localization, the results in Propositions 1, 2, 3, 4 and Corollary~1 can be considered to be valid for generic system and jamming models.}

\section{Numerical Examples}\label{sec:Simu}

In this section, the theoretical results in Section~\ref{sec:OptJamPlace} are illustrated via numerical examples. The parameters in \eqref{eq:OptProblem2} are set to $\varepsilon=1\,$m., $N_0=2$, $\nu=2$, and $K_i=1$ for $i=1,\ldots,N_T$, and the jammer power $P_J$ is normalized as $\bar{P}_J=2P_J/N_0$. For each target node, LOS connections to all the anchor nodes are assumed, and $R_i$ in \eqref{eq:OptProblem2} is calculated via \eqref{eq:ri} based on \eqref{eq:phiVec} and the following expression: $\lambda_{ij}=100\|\bs{x}_i-\bs{y}_j\|^{-2}$; that is, the free space propagation model is considered as in \cite{Moe_robust_power_allocation_2013}.

\begin{figure}
\vspace{-0.1cm}
\center
\psfrag{xlabel}[cc][][1]{horizontal [m]}
 \psfrag{ylabel}[cc][][1]{vertical [m]}
 \psfrag{t1}[cc][][.7]{Target 1}
 \psfrag{t2}[cc][][.7]{Target 2}
 \psfrag{t3}[cc][][.7]{Target 3}
 \psfrag{t4}[cc][][.7]{Target 4}
  \includegraphics[width=95mm]{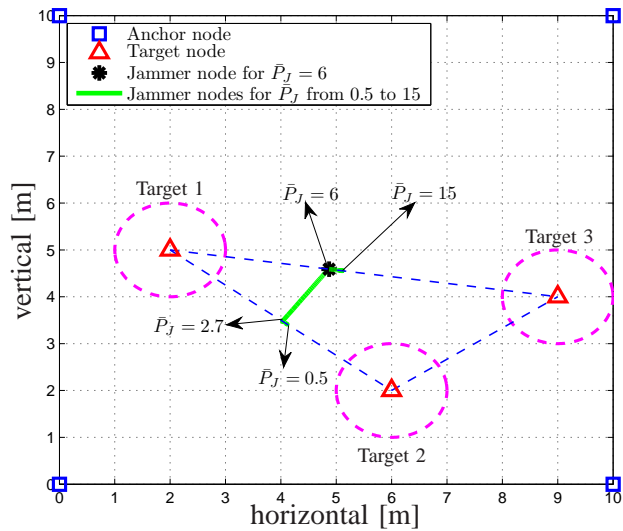}
  \vspace{-0.5cm}
\caption{The network consisting of anchor nodes at $[0~0]$, $[10~0]$, $[0~10]$, and $[10~10]\,$m., and target nodes at $[2~5]$, $[6~2]$, and $[9~4]$\,m.}
\label{fig:net3T}
\vspace{-0.2cm}
\end{figure}

\begin{figure}
\vspace{-0.2cm}
\center
\psfrag{xlabel}[cc][][1]{Normalized jammer power $\bar{P}_J$}
\psfrag{ylabel}[cc][][1]{CRLB [$\text{m}^2$]}
  \includegraphics[width=95mm]{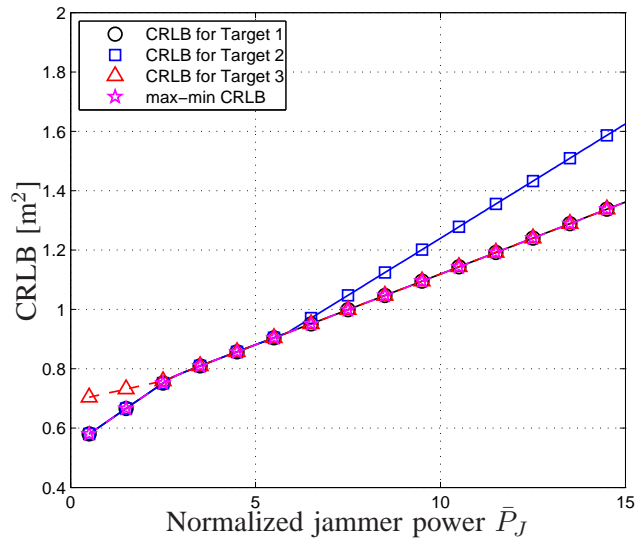}
\vspace{-0.5cm}
\caption{CRLB corresponding to each target node and max-min CRLB for the whole network for the scenario in Fig.~\ref{fig:net3T}.}
 \label{fig:net3TP}
 \vspace{-0.4cm}
\end{figure}

First, a network consisting of four anchor nodes ($N_A=4$) and three target nodes ($N_T=3$) is investigated, where the node locations are as illustrated in Fig.~\ref{fig:net3T}. For this scenario, when $\bar{P}_J=6$, {Proposition~2 can be applied as follows: $\CRLB_{\ell_1,\ell_2}$'s are calculated from \eqref{eq:OptProblem_ij}, and $\CRLB_{k,i}$ with $k=1$ and $i=3$ is found to be the minimum one. Then, it is shown that the conditions in Proposition~2 are satisfied for $k=1$ and $i=3$,} which means that the solution of the whole network (i.e., the solution of \eqref{eq:OptProblem2}) is determined by the subnetwork consisting of target node~1 and target node~3. Then, {Proposition~3 is invoked, and the optimal location of the jammer node and the corresponding max-min CRLB are calculated as $\bs{z}_{1,3}^{\rm{opt}}=[4.8713~4.5898]\,$m. and $\CRLB_{1,3}=0.9279\,{\text{m}}^2$, respectively, based on Proposition~3\textit{-(ii)-(c)}.}
In Fig.~\ref{fig:net3T}, the optimal locations of the jammer node are also shown (via the green line) for various values of $\bar{P}_J$ ranging from $0.5$ to $15$. In this scenario, the condition in {Proposition~6-(b)} is satisfied for $\ell_1=1$ and $\ell_2=2$ when $\bar{P}_J$ is lower than $2.7$, and for $\ell_1=1$ and $\ell_2=3$ when $\bar{P}_J$ is higher than $5.8$, which imply that the optimal jammer location is determined by target nodes 1 and 2 for $\bar{P}_J<2.7$, and by target nodes 1 and 3 for $\bar{P}_J>5.8$, as described in {Proposition~6-(a)}. For the remaining values of $\bar{P}_J$, the condition in {Proposition~6-(b)} is not satisfied, which implies that the solution belongs to the interior of the triangle formed by the locations of all the target nodes and that the CRLBs for all the target nodes are equalized as a result of Proposition~5. It should be noted that since the distances between the target nodes and the optimal locations of the jammer node are larger than $\varepsilon=1\,$m. (that is, the constraints in \eqref{eq:OptProblem2} are ineffective), the solution of \eqref{eq:OptProblem2} is equivalent to that obtained in the absence of the distance constraints{; hence, the results in Propositions 4-7 can be invoked.} In Fig.~\ref{fig:net3TP}, individual CRLBs of all the target nodes and the max-min CRLB of the whole network are plotted versus the normalized jammer power. From the figure, it is observed that the max-min CRLB of the whole network is equal to the CRLBs of target nodes 1 and 2 for $\bar{P}_J<2.7$, and is equal to the CRLBs of target nodes 1 and 3 for $\bar{P}_J>5.8$ in accordance with Proposition~6. For the other values of $\bar{P}_J$, the CRLBs of all the target nodes are equalized in accordance with Proposition~5 and Proposition~6.

\begin{figure}
\vspace{-0.25cm}
\center
\psfrag{xlabel}[cc][][1]{horizontal [m]}
\psfrag{ylabel}[cc][][1]{vertical [m]}
\psfrag{t1}[cc][][.7]{Target 1}
 \psfrag{t2}[cc][][.7]{Target 2}
 \psfrag{t3}[cc][][.7]{Target 3}
 \psfrag{t4}[cc][][.7]{Target 4}
  \includegraphics[width=95mm]{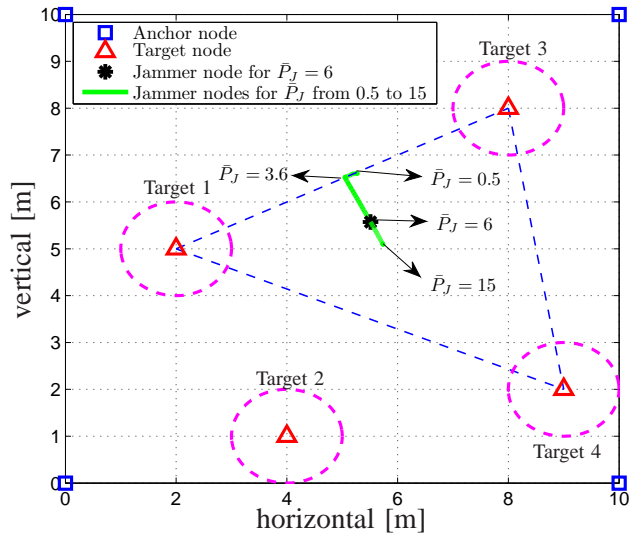}
\vspace{-0.5cm}
\caption{The network consisting of anchor nodes at $[0~0]$, $[10~0]$, $[0~10]$, and $[10~10]\,$m., and target nodes at $[2~5]$, $[4~1]$, $[8~8]$, and $[9~2]\,$m.}
 \label{fig:net4T}
 \vspace{-0.35cm}
\end{figure}

\begin{figure}
\center
\psfrag{xlabel}[cc][][1]{horizontal [m]}
\psfrag{ylabel}[cc][][1]{vertical [m]}
\psfrag{t1}[cc][][.7]{Target 1}
 \psfrag{t2}[cc][][.7]{Target 2}
 \psfrag{t3}[cc][][.7]{Target 3}
 \psfrag{t4}[cc][][.7]{Target 4}
  \includegraphics[width=95mm]{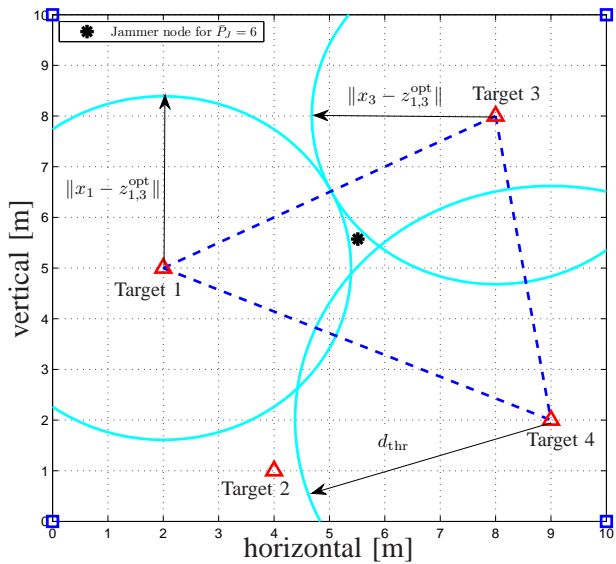}
\vspace{-0.55cm}
\caption{Illustration of Corollary 2 for the scenario in Fig.~\ref{fig:net4T}.}
 \label{fig:net4TC}
 \vspace{-0.35cm}
\end{figure}

\begin{figure}
\center
\psfrag{xlabel}[cc][][1]{Normalized jammer power $\bar{P}_J$}
\psfrag{ylabel}[cc][][1]{CRLB [$\text{m}^2$]}
  \includegraphics[width=95mm]{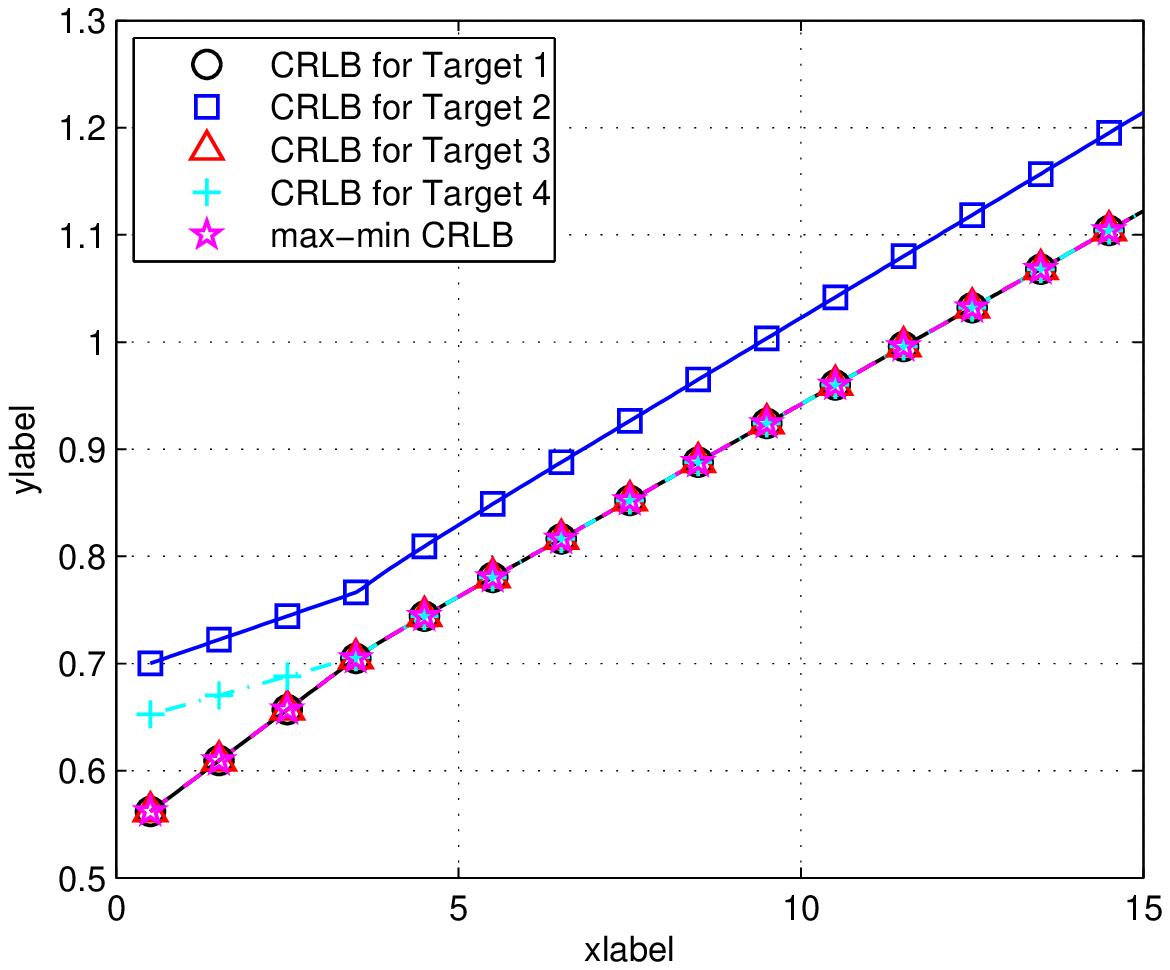}
\vspace{-0.5cm}
\caption{CRLB corresponding to each target node and max-min CRLB for the whole network for the scenario in Fig.~\ref{fig:net4T}.}
 \label{fig:net4TP}
 \vspace{-0.3cm}
\end{figure}

Next, another scenario with four anchor nodes and four target nodes is investigated, where the node locations are as shown in Fig.~\ref{fig:net4T}. {For $\bar{P}_J=6$, when Proposition~7 is employed in this scenario,} it is observed that the subnetwork consisting of target nodes 1, 3, and 4 achieves the minimum max-min CRLB among all possible subnetworks with three target nodes. In addition, the condition in {Proposition~6-(b)} is not satisfied, which implies that $\bs{z}^{\rm{opt}}_{1,3,4}$ belongs to the interior of the convex
hull (triangle) formed by the locations of target nodes 1, 3, and 4; hence, as stated by Proposition~5, the CRLBs of target nodes 1, 3, and 4 are equalized. Accordingly, the corresponding values are obtained as $\CRLB_{1,3,4}=0.7983\,{\text{m}}^2$ and $\bs{z}^{\rm{opt}}_{1,3,4}=[5.5115~5.5717]\,$m., and the calculations show that the CRLB for target node~2 is larger than $\CRLB_{1,3,4}$ for the optimal jammer location. Also, {according to Corollary~2}, the optimal location of the jammer node cannot be inside any of the circles centered at target nodes $1$, $3$, and $4$ with radii $\|\bs{x}_1-\bs{z}_{1,3}^{\rm{opt}}\|$, $\|\bs{x}_3-\bs{z}_{1,3}^{\rm{opt}}\|$, and $d_{\rm{thr}}$, respectively, which is confirmed by Fig.~\ref{fig:net4TC}. {Hence, Corollary~2 can be useful for reducing the search space for the optimal location of the jammer node.} Since the distances between the target nodes and $\bs{z}^{\rm{opt}}_{1,3,4}$ are larger than $\varepsilon=1\,$m.; that is, $\bs{z}^{\rm{opt}}_{1,3,4}$ is an element of $\{\bs{z}\,|\,\|\bs{z}-\bs{x}_i\|\geq \varepsilon\,,~i=1,2,3,4\}$, the solution of \eqref{eq:OptProblem2} is the same as that of the subnetwork consisting of target nodes 1, 3, and 4 in this scenario. In Fig.~\ref{fig:net4T}, the optimal location of the jammer node is also investigated for the values of $\bar{P}_J$ ranging from $0.5$ to $15$ (the green line in the figure). Proposition~7 indicates that the subnetwork consisting of target nodes 1, 3, and 4 achieves the minimum max-min CRLB among all possible subnetworks with three target nodes for all values of $\bar{P}_J$ in this range. It is also observed that the condition in part~(b) of Proposition~6 is satisfied with $\ell_1=1$ and $\ell_2=3$ for the values of $\bar{P}_J$ lower than $3.6$, which implies that the solution is determined by target nodes 1 and 3 for $\bar{P}_J<3.6$ as specified by part~(a) of Proposition~6. For the other values of $\bar{P}_J$, the condition in {Proposition~6-(b)} is not satisfied, indicating that the solution belongs to the interior of the triangle formed by the locations of target nodes 1, 3, and 4, and the CRLBs of target nodes 1, 3, and 4 are equalized in accordance with Proposition 5. In Fig.~\ref{fig:net4TP}, the CRLBs of the target nodes and the max-min CRLB of the whole network are plotted versus the normalized jammer power for the values of $\bar{P}_J$ ranging from $0.5$ to $15$. In accordance with the previous findings, based on Proposition 5, Proposition 6, and Proposition 7, the CRLBs of target nodes 1 and 3 are equalized to the max-min CRLB of the whole network when $\bar{P}_J$ is lower than $3.6$, and for the other values of $\bar{P}_J$ the CRLBs of target nodes 1, 3, and 4 are equalized to the max-min CRLB of the whole network.

\begin{figure}
\vspace{-0.2cm}
\center
\psfrag{xlabel}[cc][][1]{horizontal [m]}
\psfrag{ylabel}[cc][][1]{vertical [m]}
\psfrag{t1}[cc][][.7]{Target 1}
 \psfrag{t2}[cc][][.7]{Target 2}
 \psfrag{t3}[cc][][.7]{Target 3}
 \psfrag{t4}[cc][][.7]{Target 4}
 \psfrag{t5}[cc][][.7]{Target 5}
  \includegraphics[width=95mm]{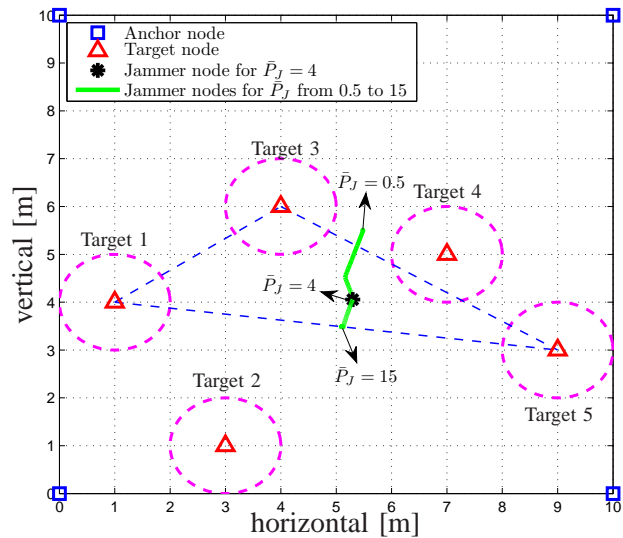}
\vspace{-0.5cm}
\caption{The network consisting of anchor nodes at $[0~0]$, $[10~0]$, $[0~10]$, and $[10~10]\,$m., and target nodes at $[1~4]$, $[3~1]$, $[4~6]$, $[7~5]$, and $[9~3]\,$m.}
 \label{fig:net5T}
 \vspace{-0.4cm}
\end{figure}


\begin{figure}
\center
\psfrag{xlabel}[cc][][1]{Normalized jammer power $\bar{P}_J$}
\psfrag{ylabel}[cc][][1]{CRLB [$\text{m}^2$]}
  \includegraphics[width=95mm]{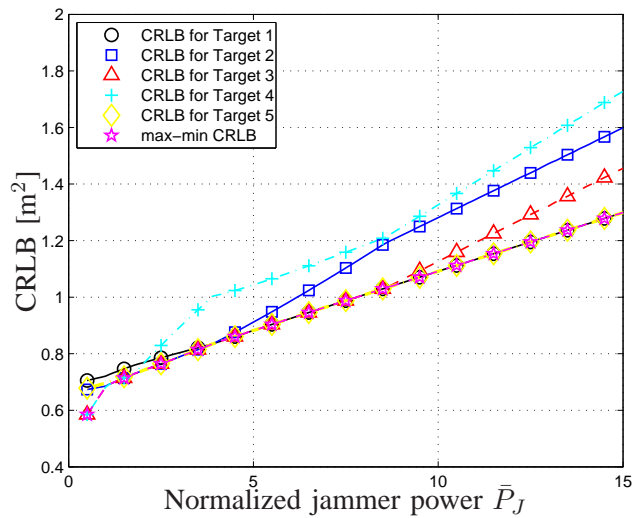}
\vspace{-0.5cm}
\caption{CRLB corresponding to each target node and max-min CRLB for the whole network for the scenario in Fig.~\ref{fig:net5T}.}
 \label{fig:net5TP}
 \vspace{-0.4cm}
\end{figure}

In the final scenario, the network in Fig.~\ref{fig:net5T} with four anchor nodes and five target nodes is considered. {Via Proposition~7,} it is calculated for $\bar{P}_J=4$ that the subnetwork consisting of target nodes 1, 3, and 5 achieves the minimum max-min CRLB among all possible subnetworks with three target nodes, and by checking the condition in {Proposition~6-(b)}, it is shown that $\bs{z}^{\rm{opt}}_{1,3,5}$ belongs to the interior of the convex hull (triangle) formed by the locations of target nodes 1, 3, and 5, and the CRLBs of target nodes 1, 3, and 5 are equalized {in compliance with Proposition 5 (see the algorithm at the end of Section~IV.)}. In accordance with these findings, the corresponding values are obtained as $\CRLB_{1,3,5}=0.8392\,{\text{m}}^2$ and $\bs{z}^{\rm{opt}}_{1,3,5}=[5.2987~4.0537]\,$m., and the CRLBs for the other target nodes are shown to be larger than $\CRLB_{1,3,5}$ for the optimal jammer location. 
In this scenario, similar to the previous scenarios, $\bs{z}^{\rm{opt}}_{1,3,5}$ is an element of $\{\bs{z}\,|\,\|\bs{z}-\bs{x}_i\|\geq \varepsilon\,,~i=1,2,3,4,5\}$; hence, the solution of \eqref{eq:OptProblem2} is the same as that of the subnetwork consisting of target nodes 1, 3, and 5.
Corollary~2 imposes that the optimal location of the jammer node cannot be inside any of the circles centered at target nodes $1$, $3$, and $5$ with radii $\|\bs{x}_1-\bs{z}_{1,5}^{\rm{opt}}\|$, $d_{\rm{thr}}$, and $\|\bs{x}_5-\bs{z}_{1,5}^{\rm{opt}}\|$, respectively,
{which can easily be verified in this example.}
In Fig.~\ref{fig:net5T}, the optimal location of the jammer node is also shown for the values of $\bar{P}_J$ ranging from $0.5$ to $15$. {In compliance with Proposition~7,} the subnetwork consisting of target nodes 2, 3, and 4 achieves the minimum max-min CRLB among all possible subnetworks with three target nodes for the values of $\bar{P}_J$ lower than $1.7$, the subnetwork consisting of target nodes 2, 3, and 5 achieves the minimum max-min CRLB for $\bar{P}_J$  between $1.7$ and $3.9$, and the subnetwork consisting of target nodes 1, 3, and 5 achieves the minimum max-min CRLB for $\bar{P}_J$ above $3.9$. Since the distances between the target nodes and the optimal location of the jammer node are larger than $\varepsilon=1\,$m. for all $\bar{P}_J$ in this scenario, the solution of \eqref{eq:OptProblem2} is the same as those of the aforementioned subnetworks for the respective ranges of $\bar{P}_J$.
Considering the values of $\bar{P}_J$ lower than $1.7$, the condition in {Proposition~6-(b)} is satisfied with $\ell_1=3$ and $\ell_2=4$ for $\bar{P}_J<1.1$, which implies that the solution is determined by target nodes 3 and 4 for $\bar{P}_J<1.1$ as described in {Proposition~6-(a)}, and for $1.1\le\bar{P}_J<1.7$ by {Proposition~6-(b)} the optimal jammer location is shown to belong to the interior of the triangle formed by the locations of target nodes 2, 3, and 4, and the CRLBs of target nodes 2, 3, and 4 are equalized {due to Proposition 5}. Similarly, based on Propositions 5 and 6, it can be shown for $1.7\leq\bar{P}_J\le3.9$  that the optimal jammer location belongs to the interior of the triangle formed by the locations of the target nodes 2, 3, and 5, and that the CRLBs of target nodes 2, 3, and 5 are equalized. In a similar fashion, it can be shown for $\bar{P}_J>3.9$ that the optimal location of the jammer node is determined only by target nodes 1 and 5 for $\bar{P}_J\ge 8.5$ as described in {Proposition 6-(a)}, and for $3.9<\bar{P}_J<8.5$ it belongs to the interior of the triangle formed by the locations of target nodes 1, 3, and 5, which results in the equalization of the CRLBs of target nodes 1, 3, and 5. In Fig.~\ref{fig:net5TP}, the CRLBs of all the target nodes and the max-min CRLB of the whole network are plotted versus the normalized jammer power for the values of $\bar{P}_J$ ranging from $0.5$ to $15$. All the previous findings are confirmed by this figure.

To analyze the effects of the SNR on the jamming performance, the max-min CRLBs for the networks in Fig.~\ref{fig:net3T}, Fig.~\ref{fig:net4T}, and Fig.~\ref{fig:net5T} are plotted in Fig.~\ref{fig:snrP} versus the spectral density level of the measurement noise, $N_0$, where $P_J=10$ is employed. As expected, an increase in $N_0$ (equivalently, a decrease in the SNR) results in a higher max-min CRLB. Since the network geometries in Fig.~\ref{fig:net3T}, Fig.~\ref{fig:net4T}, and Fig.~\ref{fig:net5T} are similar to one another (that is, in particular, the anchor nodes are located at the same positions), the max-min CRLBs for all the three networks are close to each other, as observed from Fig.~\ref{fig:snrP}. However, there also exist some variations due to the differences in the numbers and configurations of the target nodes.
\begin{figure}
\center
\psfrag{xlabel}[cc][][1]{$N_0$}
\psfrag{ylabel}[cc][][1]{max-min CRLB [$\text{m}^2$]}
  \includegraphics[width=88mm]{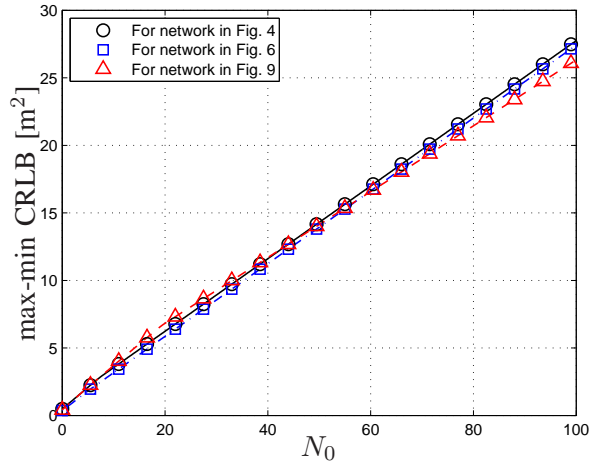}
\vspace{-0.5cm}
\caption{{Max-min CRLB for the networks in Fig.~\ref{fig:net3T}, Fig.~\ref{fig:net4T}, and Fig.~\ref{fig:net5T} versus the spectral density level of the measurement noise, $N_0$, where $P_J=10$.}}
 \label{fig:snrP}
 \vspace{-0.4cm}
\end{figure}

{For the network in Fig.~\ref{fig:net3T}, the minimum CRLB of the target nodes is plotted versus the location of the jammer node in Fig.~\ref{fig:net3TPmle}, where $N_0=2$ and $\bar{P}_J=10$ in Fig.~\ref{fig:net3TPmle}-(a) and $N_0=50$ and $\bar{P}_J=10$ in Fig.~\ref{fig:net3TPmle}-(b). In the first scenario, the optimal location of the jammer node is given by $\bs{z}_{\rm{opt}}=(5.031,4.567)\,$m. where the CRLBs of the target nodes 1 and 3 are equalized as specified by Proposition~6. On the other hand, in the second scenario, the optimal jammer location is $\bs{z}_{\rm{opt}}=(4.14,3.394)\,$m. and the CRLBs of the target nodes 1 and 2 are equalized in accordance with Proposition~6. From Fig.~\ref{fig:net3TPmle} and the location constraints shown in Fig.~4, the nonconvexity of the optimization problem in \eqref{eq:OptProblem2} can be observed clearly. In addition, it is noted that the minimum CRLB becomes more sensitive to the location of the jammer node when the spectral density level of the measurement noise is lower; that is, the minimum CRLB changes by larger factors with respect to the jammer location in Fig.~\ref{fig:net3TPmle}-(a).}

\begin{figure}
\vspace{-0.3cm}
\center
\subfigure[]{
\includegraphics[width=90mm]{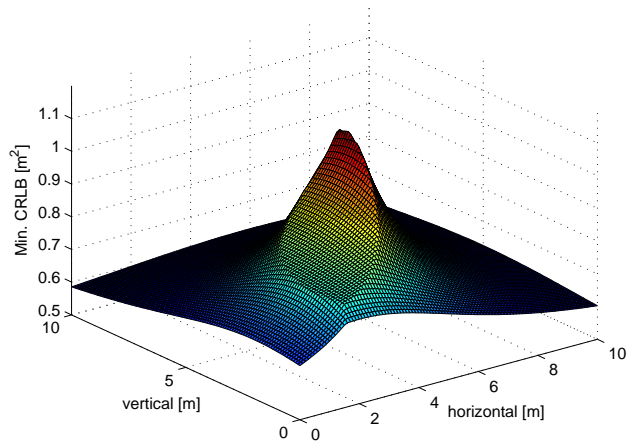}
}
\subfigure[]{
\includegraphics[width=90mm]{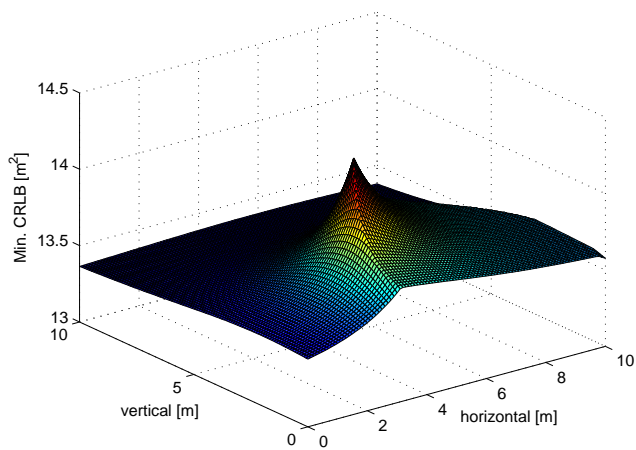}
}
\caption{{The minimum CRLB of the target nodes versus the location of the jammer node for (a) $N_0=2$ and (b) $N_0=50$, where $P_J=10$.}}
 \label{fig:net3TPmle}
\end{figure}

{In order to investigate the optimal jammer placement problem based on the CRLB expression in \eqref{eq:CRLBi_2_New} and \eqref{eq:ri_New} in Section~\ref{sec:Exten}, consider a critical SNR level for the receivers of the target nodes as $\SNRt=1$ (i.e., $0\,$dB). In addition, let the $E_{ij}$ parameter in \eqref{eq:SNRcond} be given by $E_{ij}=2000/\|\bs{x}_i-\bs{y}_j\|^2$. Then, it can be shown that the critical distances, $d_{ij}^{\rm{lim}}$, are lower than $\varepsilon=1\,$m. (cf.~\eqref{eq:OptProblem1}) in all the cases considered in the previous numerical examples. Hence, the results are valid for the CRLB expression in \eqref{eq:CRLBi_2_New} and \eqref{eq:ri_New}, as well. To provide an example in which the differences due to the CRLB expression in Section~\ref{sec:Exten} can be observed, reconsider the network in Fig.~\ref{fig:net3T} in the presence of higher powers for the jammer node. Fig.~\ref{fig:net3TPext} illustrates the CRLBs for the target nodes, together with the max-min CRLB, where the optimal locations for the jammer node are obtained based on the CRLB expression in \eqref{eq:CRLBi_2_New} and \eqref{eq:ri_New}. For comparison purposes, the max-min CRLB corresponding to the optimal locations for the jammer node obtained from the CRLB expression in \eqref{eq:CRLBi_2} and \eqref{eq:ri} is also illustrated in the figure (labeled as ``original''). It is noted that there exist discontinuities in the CRLBs due to the fact that the connections between the anchor and target nodes are lost when the SNRs get below the critical SNR level (cf.~\eqref{eq:CRLBi_2_New} and \eqref{eq:ri_New}). Also, up to $\bar{P}_J=338.5$, the max-min CRLBs with and without the consideration of lost connections take the same values. Considering that both of the max-min CRLBs achieve the value of $17.23\,{\text{m}}^2$ just before $\bar{P}_J=338.5$ and that the maximum distance between the anchor nodes is equal to $10\sqrt{2}\,\text{m}$ in the network, it can be concluded that the extended formulation based on the CRLB expression in \eqref{eq:CRLBi_2_New} and \eqref{eq:ri_New} reduces to the original formulation based on the CRLB expression in \eqref{eq:CRLBi_2} and \eqref{eq:ri} for the practical ranges of localization accuracy in this example (i.e., the differences are observed only for the cases in which the localization accuracy is unacceptable for practical applications). From Fig.~\ref{fig:net3TPext}, it is also observed that the CRLBs of target node $2$, $1$, and $3$ go to infinity at $\bar{P}_J=419.5$, $\bar{P}_J=468.6$, and $\bar{P}_J=747.1$, respectively, due to the loss of connections to the anchor nodes. As a result, the max-min CRLB becomes infinity after $\bar{P}_J=747.1$. Table~\ref{tab:jamlocext} presents the optimal jammer locations according to both formulations for various values of the normalized jammer power. It is noted that the change in the optimal location of the jammer node with respect to $\bar{P}_J$ is relatively noticeable according to the extended formulation compared to that according to the original formulation, for which the change is almost indiscernible.}

\begin{figure}
\vspace{-0.3cm}
\center
\psfrag{xlabel}[cc][][1]{Normalized jammer power $\bar{P}_J$}
\psfrag{ylabel}[cc][][1]{CRLB [$\text{m}^2$]}
  \includegraphics[width=90mm]{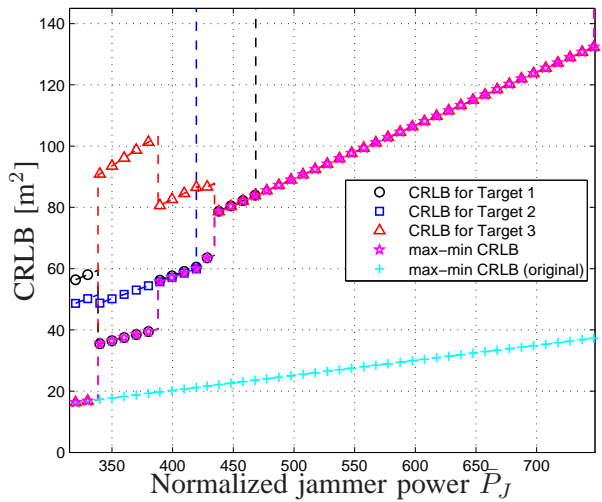}
\vspace{-0.5cm}
\caption{{CRLB of each target node and the max-min CRLB of the network for the scenario in Fig.~\ref{fig:net3T}, where the optimal locations for the jammer node are obtained based on the CRLB expression in \eqref{eq:CRLBi_2_New} and \eqref{eq:ri_New}. The max-min CRLB corresponding to the optimal locations based on the CRLB expression in \eqref{eq:CRLBi_2} and \eqref{eq:ri} is also shown (`original').}}
 \label{fig:net3TPext}
\end{figure}

\begin{table}
\center
 \caption{{The optimal location of the jammer node according to the original and extended formulations for the scenario in Fig.~\ref{fig:net3T}.}}
\begin{tabular}{|c|c|c|}
\hline
$\bar{P}_J$ & {Original formulation} & {Extended formulation}\\
 \hline
 \hline
 320  &  (5.2802, 4.5314) m. & (5.2802,  4.5314) m.  \\
 \hline
 339  & (5.2807, 4.5313) m. &  (5.4610, 4.5046) m. \\
  \hline
 420&   (5.2822, 4.5311) m. &   (4.9232, 4.7215) m. \\
 \hline
  470&   (5.2829, 4.5310) m. &  (4.6000,  4.6286) m. \\
 \hline
  747&  (5.2849, 4.5306) m. &  (4.6092, 4.6286) m. \\
 \hline
\end{tabular}
\label{tab:jamlocext}
\vspace{-0.3cm}
\end{table}

\section{Concluding Remarks}\label{sec:Conc}

The problem of optimal jammer placement has been proposed for maximizing the minimum of the CRLBs for a number of target nodes in a wireless localization system. Theoretical results have been obtained for specifying scenarios in which the jammer node is located as close to a certain target node as possible, or the optimal location of the jammer node is determined by two of the target nodes. Also, explicit expressions for the optimal location of the jammer node have been derived in the presence of two target nodes. In the absence of distance constraints for the jammer node, it has been shown that the optimal jammer location lies on the convex hull formed by the locations of the target nodes, equalizes the CRLBs of at least two of the target nodes, and is determined by two or three of the target nodes. Numerical examples have provided an illustration of the theoretical results in different scenarios. {Performing experiments to evaluate the effects of jamming and to investigate the optimal location for a jammer node in a practical wireless localization system can be considered as an important direction for future work.}


{Based on the results in this manuscript, various guidelines can be provided related to jamming mitigation in wireless localization systems. Since the solution of the optimal jammer placement problem (cf.~\eqref{eq:OptProblem2}) corresponds to the maximum degradation that can be caused by a jammer node, the transmitted powers of the anchor nodes in the wireless localization system can be adjusted accordingly in order to satisfy certain accuracy requirements in all scenarios. (A target node can measure the received noise level in certain intervals to determine the presence and power level of the jammer node.) In addition, for applications in which the anchors nodes can be moved, the locations of the anchor nodes (hence, the geometry of the system) can be adapted for reducing the effectiveness of jamming (cf.~\eqref{eq:ri}). Furthermore, if possible, additional anchor nodes can be employed depending on the required localization accuracy and the severity of jamming.}

Although the jammer node is assumed to know all the localization related parameters in this study, the results can also be extended to scenarios with certain types of uncertainty. For example, if $R_1,\ldots,R_{N_T}$, $K_1,\ldots,K_{N_T}$, and $N_0$ in \eqref{eq:OptProblem2} are confined to linear uncertainty sets as in \cite{Moe_robust_power_allocation_2013}, it can be shown that a robust jammer placement algorithm can be designed based on the minimum possible values of these parameters in the uncertainty sets. Since the structure of the CRLB expressions will not change, all the theoretical results will be valid in that scenario, as well.\footnote{In practice, the jammer node can obtain information about the localization parameters by, e.g., using cameras to learn the locations of the target and anchor nodes, performing prior measurements in the environment to form a database for the channel parameters, and listening to signals between the anchor and target nodes \cite{Gezici_JammerPowAlloc1}.}

\appendix
\section{Appendices}
\subsection{{Proof of Proposition 1}}\label{app:Prop1}

First, an upper bound is derived for the optimization problem in \eqref{eq:OptProblem2} as follows:
\begin{align}\label{eq:Prop1_uppBnd3}
\underset{\bs{z}}{\max}\,\underset{i\in\{1,\ldots,N_T\}}\min~
R_i\left(\frac{K_iP_J}{\|\bs{z}-\bs{x}_i\|^\nu}+\frac{N_0}{2}\right)
\leq\underset{\bs{z}}{\max}~R_\ell\left(\frac{K_\ell P_J}{\|\bs{z}-\bs{x}_\ell\|^\nu}+\frac{N_0}{2}\right)
=R_\ell\left(\frac{K_\ell P_J}{\varepsilon^\nu}+\frac{N_0}{2}\right)
\end{align}
where the inequality follows by definition, and the equality is obtained from the constraint in \eqref{eq:OptProblem2}.
Next, towards the aim of proving the achievability of the upper bound in \eqref{eq:Prop1_uppBnd3} under the conditions in the proposition, the following relation is presented for $i\in\{1,\ldots,N_T\}\setminus\{\ell\}$ and for all $\bs{z}$ such that $\|\bs{z}-\bs{x}_\ell\|=\varepsilon\,$:
\begin{align}\label{eq:Prop1_uppBnd4}
R_i\left(\frac{K_iP_J}{\|\bs{z}-\bs{x}_i\|^\nu}+\frac{N_0}{2}\right)
\geq
R_i\left(\frac{K_i P_J}{(\|\bs{x}_i-\bs{x}_\ell\|+\varepsilon)^\nu}+\frac{N_0}{2}\right)
\geq R_\ell\left(\frac{K_\ell P_J}{\varepsilon^\nu}+\frac{N_0}{2}\right)
\end{align}
where the first inequality follows from the triangle inequality; that is, $\|\bs{z}-\bs{x}_i\|\leq\|\bs{x}_i-\bs{x}_\ell\|+\|\bs{z}-\bs{x}_\ell\|
=\|\bs{x}_i-\bs{x}_\ell\|+\varepsilon$, and the second inequality is due to the condition in \eqref{eq:Prop1_ineq}. The inequality in \eqref{eq:Prop1_uppBnd4} for $i\in\{1,\ldots,N_T\}\setminus\{\ell\}$ implies that, for $\|\bs{z}-\bs{x}_\ell\|=\varepsilon$ and under the condition in \eqref{eq:Prop1_ineq}, the upper bound in \eqref{eq:Prop1_uppBnd3} can be achieved as follows:
$\underset{i\in\{1,\ldots,N_T\}}\min\,
R_i\left(\frac{K_iP_J}{\|\bs{z}-\bs{x}_i\|^\nu}+\frac{N_0}{2}\right)
=R_\ell\left(\frac{K_\ell P_J}{\|\bs{z}-\bs{x}_l\|^\nu}+\frac{N_0}{2}\right)
=R_\ell\left(\frac{K_\ell P_J}{\varepsilon^\nu}+\frac{N_0}{2}\right)$
if set $\{\bs{z}\,:\,\|\bs{z}-\bs{x}_\ell\|=\varepsilon~\&~\|\bs{z}-\bs{x}_i\|\geq\varepsilon,\, i=1,\ldots,\ell-1,\ell+1,\ldots,N_T\}$ is non-empty. In other words, under the conditions in the proposition, the optimization problem in \eqref{eq:OptProblem2} achieves the upper bound in \eqref{eq:Prop1_uppBnd3} for $\|\bs{z}-\bs{x}_\ell\|=\varepsilon$. Hence, the solution $\bs{z}^{\rm{opt}}$ of \eqref{eq:OptProblem2} satisfies $\|\bs{z}^{\rm{opt}}-\bs{x}_\ell\|=\varepsilon$ if the conditions in the proposition hold.\hfill$\blacksquare$

\subsection{{Proof of Proposition 3}}\label{app:Prop3}

\textit{\textbf{(i)}} If $\|\bs{x}_1-\bs{x}_2\| < 2\,\varepsilon$, the optimal location for the jammer node, $\bs{z}^{\rm{opt}}$, is equal to one of the two intersection points of the circles centered at $\bs{x}_1$ and $\bs{x}_2$ with radii $\varepsilon$. In that case, $\|\bs{z}^{\rm{opt}}-\bs{x}_1\|=\|\bs{z}^{\rm{opt}}-\bs{x}_2\|=\varepsilon$ is obtained, which achieves the upper bound of the problem in \eqref{eq:OptProblem2} for $N_T=2$. Hence, the solution of \eqref{eq:OptProblem2} is given by $\bs{z}^{\rm{opt}}$.

\textit{\textbf{(ii)}} Suppose that $\|\bs{x}_1-\bs{x}_2\| \geq 2\,\varepsilon$. Consider the straight line segment between $\bs{x}_1$ and $\bs{x}_2$. Let $\bs{z}_1$ and $\bs{z}_2$ denote, respectively, the intersections of this line segment with the circles centered at $\bs{x}_1$ and $\bs{x}_2$ with radii $\varepsilon$, as illustrated in Fig.~\ref{fig:prop3}.
\begin{figure}
\vspace{-0.1cm}
\center
  \includegraphics[width=80mm]{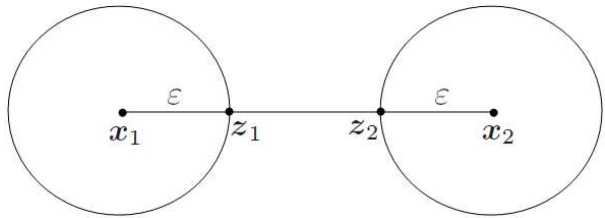}
\vspace{-0.2cm}
\caption{Illustration of the scenario in Part \textit{(ii)} of Proposition~3.}
 \label{fig:prop3}
 \vspace{-0.3cm}
\end{figure}
Denote the straight line segment between $\bs{z}_1$ and $\bs{z}_2$ as $L_{12}$. First, it can be proved that for any feasible location $\bs{z}^+$ that is not on $L_{12}$, there exists a location $\bs{z}^*$ on $L_{12}$ which satisfies either $\|\bs{z}^*-\bs{x}_1\|<\|\bs{z}^+-\bs{x}_1\|$ $\&$ $\|\bs{z}^*-\bs{x}_2\|\leq\|\bs{z}^+-\bs{x}_2\|$ or $\|\bs{z}^*-\bs{x}_1\|\leq\|\bs{z}^+-\bs{x}_1\|$ $\&$  $\|\bs{z}^*-\bs{x}_2\|<\|\bs{z}^+-\bs{x}_2\|$ (the detailed proof for this statement is not included due to the space limitation).
Since, the CRLB is inversely proportional to the distance between the jammer and target nodes, it is concluded that $\bs{z}^+$ (i.e., any location not on $L_{12}$) cannot be a solution of \eqref{eq:OptProblem2} for $N_T=2$. Hence, the optimal location for the jammer node must satisfy $\|\bs{z}^{\rm{opt}}-\bs{x}_1\|+\|\bs{z}^{\rm{opt}}-\bs{x}_2\|=\|\bs{x}_2-\bs{x}_1\|$ together with the distance constraints $\|\bs{z}^{\rm{opt}}-\bs{x}_1\|\geq\varepsilon$ and $\|\bs{z}^{\rm{opt}}-\bs{x}_2\|\geq\varepsilon$. In addition, if the condition in $(ii)$--$(a)$ is satisfied, it means that the CRLB for target node $1$ is the minimum CRLB for all $\bs{z}$ on $L_{12}$; hence, the optimal solution is to place the jammer node as close to target node 1 as possible in this case; i.e., $\|\bs{z}^{\rm{opt}}-\bs{x}_1\|=\varepsilon$. Similarly, if the condition in $(ii)$--$(b)$ is satisfied, the CRLB for target node $2$ becomes the minimum CRLB for all $\bs{z}$ on $L_{12}$, and $\|\bs{z}^{\rm{opt}}-\bs{x}_2\|=\varepsilon$ is obtained. For the condition in $(ii)$--$(c)$, first suppose that $\|\bs{z}^{\rm{opt}}-\bs{x}_1\|>d^*$, where $d^*$ is as defined in the proposition. In this case, the CRLB for target node 1 becomes the minimum, which is lower than $R_1({K_1P_J}/{(d^*)^{\,\nu}}+{N_0}/{2})$ (see \eqref{eq:equalizer2}). Hence, a contradiction arises, implying that $\|\bs{z}^{\rm{opt}}-\bs{x}_1\|>d^*$ cannot hold. Similarly, in the case of $\|\bs{z}^{\rm{opt}}-\bs{x}_1\|<d^*$, the CRLB for target node 2 becomes the minimum, which is lower than $R_2({K_2P_J}/{(\|\bs{x}_1-\bs{x}_2\|-d^*)^\nu}+{N_0}/{2})$ (see \eqref{eq:equalizer2}), which leads to a contradiction. Hence, the optimal solution must satisfy $\|\bs{z}^{\rm{opt}}-\bs{x}_1\|=d^*$ under the condition in $(ii)$--$(c)$.\hfill$\blacksquare$

\subsection{{Proof of Proposition 5}}\label{app:Prop5}

Consider target nodes $\ell_1$, $\ell_2$, and $\ell_3$, and let $\bs{z}^{\rm{opt}}_{\ell_1,\ell_2,\ell_3}$ denote the optimizer of \eqref{eq:optProb_3tar_NoConstr}. Also, let $\mathcal{H}$ represent the convex hull formed by the locations of the target nodes, which corresponds to a triangle with the target nodes at the vertices. As stated in the proposition, $\bs{z}^{\rm{opt}}_{\ell_1,\ell_2,\ell_3}$ belongs to the interior of $\mathcal{H}$. 

First, suppose that the CRLB for one of the target nodes is the minimum and those for the other target nodes are strictly larger for $\bs{z}^{\rm{opt}}_{\ell_1,\ell_2,\ell_3}$. Without loss of generality, let
$\CRLB_{\ell_1}(\bs{z}^{\rm{opt}}_{\ell_1,\ell_2,\ell_3})>
\CRLB_{\ell_3}(\bs{z}^{\rm{opt}}_{\ell_1,\ell_2,\ell_3})$ and $\CRLB_{\ell_2}(\bs{z}^{\rm{opt}}_{\ell_1,\ell_2,\ell_3})>
\CRLB_{\ell_3}(\bs{z}^{\rm{opt}}_{\ell_1,\ell_2,\ell_3})$. In this case, $\CRLB_{\ell_1,\ell_2,\ell_3}$ in \eqref{eq:optProb_3tar_NoConstr} is equal to $\CRLB_{\ell_3}(\bs{z}^{\rm{opt}}_{\ell_1,\ell_2,\ell_3})$. Then, consider the projection of $\bs{z}^{\rm{opt}}_{\ell_1,\ell_2,\ell_3}$ onto the straight line that passes through target nodes $\ell_2$ and $\ell_3$, and denote it by $\bs{z}_0$. Since $\bs{z}^{\rm{opt}}_{\ell_1,\ell_2,\ell_3}$ belongs to the interior of $\mathcal{H}$, there exists $\Delta>0$ such that $\bs{z}_\delta\triangleq\bs{z}^{\rm{opt}}_{\ell_1,\ell_2,\ell_3}+
(\bs{z}_0-\bs{z}^{\rm{opt}}_{\ell_1,\ell_2,\ell_3})\delta/
\|\bs{z}_0-\bs{z}^{\rm{opt}}_{\ell_1,\ell_2,\ell_3}\|$ belongs to the interior of $\mathcal{H}$ for $\delta\in(0,\Delta)$ (see Fig.~\ref{fig:prop5}-(a) for illustration).
\begin{figure}
\vspace{-0.1cm}
\center
  \includegraphics[width=90mm]{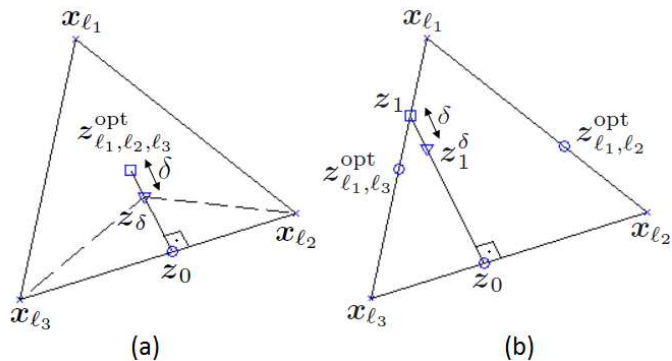}
\vspace{-0.2cm}
\caption{(a) Illustration for the proof of Proposition~5. (b) Illustration for the proof of Proposition~6.}
 \label{fig:prop5}
\vspace{-0.3cm}
\end{figure}
For a given value of $\delta\in(0,\Delta)$, $\bs{z}_\delta$ also corresponds to the projection of $\bs{z}^{\rm{opt}}_{\ell_1,\ell_2,\ell_3}$ onto the triangle with vertices at $\bs{z}_\delta$, $\bs{x}_{\ell_2}$, and $\bs{x}_{\ell_3}$. Therefore, based on similar arguments to those in Proposition~4, the projection theorem \cite{Bertsekas2} can be invoked to show that $\bs{z}_\delta$ is closer to both target node $\ell_2$ and target node $\ell_3$ than $\bs{z}^{\rm{opt}}_{\ell_1,\ell_2,\ell_3}$; that is,
\begin{align}\label{eq:ineqEqu1}
\|\bs{z}_\delta-\bs{x}_{\ell_2}\|&<\|\bs{z}^{\rm{opt}}_{\ell_1,\ell_2,\ell_3}-\bs{x}_{\ell_2}\|\\\label{eq:ineqEqu2} \|\bs{z}_\delta-\bs{x}_{\ell_3}\|&<\|\bs{z}^{\rm{opt}}_{\ell_1,\ell_2,\ell_3}-\bs{x}_{\ell_3}\|\,. \end{align}
Based on Lemma~1 in Appendix~\ref{app:ForProp5}, \eqref{eq:ineqEqu1} and \eqref{eq:ineqEqu2} implies that
\begin{gather}\label{eq:ineqEqu3}
\|\bs{z}_\delta-\bs{x}_{\ell_1}\|\geq\|\bs{z}^{\rm{opt}}_{\ell_1,\ell_2,\ell_3}-\bs{x}_{\ell_1}\|\,.
\end{gather}
From \eqref{eq:ineqEqu1}--\eqref{eq:ineqEqu3}, it is concluded via \eqref{eq:CRLBm} that
\begin{align}\label{eq:ineqEqu4}
\CRLB_{\ell_1}(\bs{z}^{\rm{opt}}_{\ell_1,\ell_2,\ell_3})&\geq\CRLB_{\ell_1}(\bs{z}_\delta)\,,\\\label{eq:ineqEqu5} \CRLB_{\ell_2}(\bs{z}^{\rm{opt}}_{\ell_1,\ell_2,\ell_3})&<\CRLB_{\ell_2}(\bs{z}_\delta)\,, \\\label{eq:ineqEqu6} \CRLB_{\ell_3}(\bs{z}^{\rm{opt}}_{\ell_1,\ell_2,\ell_3})&<\CRLB_{\ell_3}(\bs{z}_\delta)\,.
\end{align}
Since $\CRLB_{\ell_1}(\bs{z}^{\rm{opt}}_{\ell_1,\ell_2,\ell_3})>\CRLB_{\ell_3}(\bs{z}^{\rm{opt}}_{\ell_1,\ell_2,\ell_3})$ and the CRLB in \eqref{eq:CRLBm} is a continuous function of the distance, there exists $\delta\in(0,\Delta)$ such that
\begin{align}\label{eq:ineqEqu7}
\CRLB_{\ell_1}(\bs{z}_\delta)>\CRLB_{\ell_3}(\bs{z}^{\rm{opt}}_{\ell_1,\ell_2,\ell_3})
=\CRLB_{\ell_1,\ell_2,\ell_3}(\bs{z}^{\rm{opt}}_{\ell_1,\ell_2,\ell_3})\,.
\end{align}
The relations in \eqref{eq:ineqEqu4}--\eqref{eq:ineqEqu7} together with
$\CRLB_{\ell_1}(\bs{z}^{\rm{opt}}_{\ell_1,\ell_2,\ell_3})>
\CRLB_{\ell_3}(\bs{z}^{\rm{opt}}_{\ell_1,\ell_2,\ell_3})$
and
$\CRLB_{\ell_2}(\bs{z}^{\rm{opt}}_{\ell_1,\ell_2,\ell_3})>
\CRLB_{\ell_3}(\bs{z}^{\rm{opt}}_{\ell_1,\ell_2,\ell_3})$ imply that there exists $\delta\in(0,\Delta)$ such that $\CRLB_{\ell_1,\ell_2,\ell_3}(\bs{z}_\delta)>
\CRLB_{\ell_1,\ell_2,\ell_3}(\bs{z}^{\rm{opt}}_{\ell_1,\ell_2,\ell_3})$. Therefore, $\bs{z}^{\rm{opt}}_{\ell_1,\ell_2,\ell_3}$ is not optimal, which leads to a contradiction. Hence, it is not possible that the CRLB for one of the target nodes is the minimum and those for the other target nodes are strictly larger for $\bs{z}^{\rm{opt}}_{\ell_1,\ell_2,\ell_3}$.

Secondly, suppose that two of the CRLBs for the target nodes are the same and that for the other target node is larger. Without loss of generality, let $\CRLB_{\ell_1}(\bs{z}^{\rm{opt}}_{\ell_1,\ell_2,\ell_3})>
\CRLB_{\ell_2}(\bs{z}^{\rm{opt}}_{\ell_1,\ell_2,\ell_3})=
\CRLB_{\ell_3}(\bs{z}^{\rm{opt}}_{\ell_1,\ell_2,\ell_3})$. Based on the same arguments as in the previous case, it can be shown that there exists $\bs{z}_\delta$ for which the relations in \eqref{eq:ineqEqu4}--\eqref{eq:ineqEqu7} hold. Therefore, $\CRLB_{\ell_1,\ell_2,\ell_3}(\bs{z}_\delta)>
\CRLB_{\ell_1,\ell_2,\ell_3}(\bs{z}^{\rm{opt}}_{\ell_1,\ell_2,\ell_3})$ is obtained, resulting in a contradiction. Hence, the only feasible scenario in which $\bs{z}^{\rm{opt}}_{\ell_1,\ell_2,\ell_3}$ belongs to the interior of $\mathcal{H}$ is the one with $\CRLB_{\ell_1}(\bs{z}^{\rm{opt}}_{\ell_1,\ell_2,\ell_3})=
\CRLB_{\ell_2}(\bs{z}^{\rm{opt}}_{\ell_1,\ell_2,\ell_3})=
\CRLB_{\ell_3}(\bs{z}^{\rm{opt}}_{\ell_1,\ell_2,\ell_3})$.\hfill$\blacksquare$


\subsection{An Auxiliary Result}\label{app:ForProp5}

\textit{\textbf{Lemma 1:} Consider a triangle in a two-dimensional space with vertices $A$, $B$, and $C$, and a point $P_1$ inside the triangle. Let $d_{A,1}$, $d_{B,1}$, and $d_{C,1}$ denote the distances of $P_1$ from vertices $A$, $B$, and $C$, respectively. Consider another point $P_2$ on the triangle with distances $d_{A,2}$, $d_{B,2}$, and $d_{C,2}$ from vertices $A$, $B$, and $C$, respectively. If $d_{B,2}\leq d_{B,1}$ and $d_{C,2}\leq d_{C,1}$, then $d_{A,2}\geq d_{A,1}$.}

{The proof is not presented due to the space limitation.}

\subsection{{Proof of Proposition 6}}\label{app:Prop6}

{\textbf{Part a)}:} Consider the scenario in which the optimal jammer location is on the boundary of the triangle formed by target nodes $\ell_1$, $\ell_2$, and $\ell_3$. First, suppose that $\bs{z}_1$ is an optimal location for the jammer node, which lies on the straight line segment between $\bs{x}_{\ell_1}$ and $\bs{z}_{\ell_1,\ell_3}^{\rm{opt}}$, where $\bs{x}_{\ell_1}$ is the location of target node $\ell_1$ and $\bs{z}_{\ell_1,\ell_3}^{\rm{opt}}$ is the optimizer of \eqref{eq:optProb_2tar_NoConstr} for target nodes $\ell_1$ and $\ell_3$, which corresponds to $\CRLB_{\ell_1,\ell_3}$. As stated in the proposition, $\CRLB_{\ell_1,\ell_2}<\CRLB_{\ell_1,\ell_3}$. Therefore, due to the equalizer property in Corollary~1, $\CRLB_{\ell_1}(\bs{z}_{\ell_1,\ell_2}^{\rm{opt}})<\CRLB_{\ell_1}(\bs{z}_{\ell_1,\ell_3}^{\rm{opt}})$ must hold. Then, the following relations are obtained:
\begin{gather}\label{eq:distRelProp6_1}
\|\bs{x}_{\ell_1}-\bs{z}_{\ell_1,\ell_2}^{\rm{opt}}\|
>\|\bs{x}_{\ell_1}-\bs{z}_{\ell_1,\ell_3}^{\rm{opt}}\|
\geq\|\bs{x}_{\ell_1}-\bs{z}_1\|
\end{gather}
where the first inequality follows from $\CRLB_{\ell_1}(\bs{z}_{\ell_1,\ell_2}^{\rm{opt}})<\CRLB_{\ell_1}(\bs{z}_{\ell_1,\ell_3}^{\rm{opt}})$ and \eqref{eq:CRLBm}, and the second inequality is by the definition of location $\bs{z}_1$ (see Fig.~\ref{fig:prop5}-(b) for illustration).
The inequality in \eqref{eq:distRelProp6_1} and the equalizer property in Corollary~1 imply that
\begin{gather}\label{eq:distRelProp6_2}
\CRLB_{\ell_1}(\bs{z}_1)>\CRLB_{\ell_1}(\bs{z}_{\ell_1,\ell_2}^{\rm{opt}})=\CRLB_{\ell_1,\ell_2}\,.
\end{gather}
On the other hand, due to the definitions in \eqref{eq:optProb_3tar_NoConstr} and \eqref{eq:optProb_2tar_NoConstr}, the following relation always holds:
\begin{gather}\label{eq:distRelProp6_3}
\CRLB_{\ell_1,\ell_2,\ell_3}\leq\CRLB_{\ell_1,\ell_2}\,.
\end{gather}
Since $\bs{z}_1$ is an optimal solution of \eqref{eq:optProb_3tar_NoConstr}, $\CRLB_{\ell_1,\ell_2,\ell_3}$ is equal to $\min\{\CRLB_{\ell_1}(\bs{z}_1),\CRLB_{\ell_2}(\bs{z}_1),\CRLB_{\ell_3}(\bs{z}_1)\}$, which, together with \eqref{eq:distRelProp6_2} and \eqref{eq:distRelProp6_3}, imply that $\CRLB_{\ell_1}(\bs{z}_1)$ is not a minimum of $\{\CRLB_{\ell_1}(\bs{z}_1),\CRLB_{\ell_2}(\bs{z}_1),\CRLB_{\ell_3}(\bs{z}_1)\}$. Then, a new location $\bs{z}_1^\delta$ is defined, which is at distance of $\delta>0$ from $\bs{z}_1$ and is on the straight line segment between $\bs{z}_1$ and the projection of $\bs{z}_1$ on the straight line that passes through $\bs{x}_{\ell_2}$ and $\bs{x}_{\ell_3}$, as shown in Fig.~\ref{fig:prop5}-(b).\footnote{Note that $\bs{z}_1^\delta$ is not required to be on the triangle formed by the locations of the target nodes; it may also be outside that triangle.} 
Since the distance between $\bs{z}_1^\delta$ and $\bs{x}_{\ell_2}$ ($\bs{x}_{\ell_3}$) is smaller than the distance between $\bs{z}_1$ and $\bs{x}_{\ell_2}$ ($\bs{x}_{\ell_3}$)
(based on the projection theorem \cite{Bertsekas2} and similar arguments to those in the proof of Proposition~4), the following relations are obtained from \eqref{eq:CRLBm}:
\begin{align}\label{eq:distRelProp6_4}
\CRLB_{\ell_2}(\bs{z}_1^\delta)&>\CRLB_{\ell_2}(\bs{z}_1)\\\label{eq:distRelProp6_5} \CRLB_{\ell_3}(\bs{z}_1^\delta)&>\CRLB_{\ell_3}(\bs{z}_1)
\end{align}
In addition, since the CRLB is a continuous function of the distance, there always exists a sufficiently small $\delta>0$ such that $\CRLB_{\ell_1}(\bs{z}_1^\delta)>\CRLB_{\ell_1,\ell_2}$ (see \eqref{eq:distRelProp6_2}). Hence, based on similar arguments to those above, $\CRLB_{\ell_1}(\bs{z}_1^\delta)$ is not the minimum of $\{\CRLB_{\ell_1}(\bs{z}_1^\delta),\CRLB_{\ell_2}(\bs{z}_1^\delta),\CRLB_{\ell_3}(\bs{z}_1^\delta)\}$. Therefore, based on \eqref{eq:distRelProp6_4} and \eqref{eq:distRelProp6_5}, it is concluded that
\begin{align}\label{eq:distRelProp6_6}
\min\{\CRLB_{\ell_1}(\bs{z}_1^\delta),\CRLB_{\ell_2}(\bs{z}_1^\delta),\CRLB_{\ell_3}(\bs{z}_1^\delta)\}
>\min\{\CRLB_{\ell_1}(\bs{z}_1),\CRLB_{\ell_2}(\bs{z}_1),\CRLB_{\ell_3}(\bs{z}_1)\}
\end{align}
which contradicts the optimality of $\bs{z}_1$. Hence, it is proved via contradiction that no locations on the straight line segment between $\bs{x}_{\ell_1}$ and $\bs{z}_{\ell_1,\ell_3}^{\rm{opt}}$ can be optimal.

Secondly, suppose that $\bs{z}_2$ is an optimal location for the jammer node, which lies on the straight line segment between $\bs{x}_{\ell_3}$ and $\bs{z}_{\ell_1,\ell_3}^{\rm{opt}}$, where $\bs{z}_{\ell_1,\ell_3}^{\rm{opt}}$ is the optimizer of \eqref{eq:optProb_2tar_NoConstr} for target nodes $\ell_1$ and $\ell_3$, corresponding to $\CRLB_{\ell_1,\ell_3}$. Let the upper bound in \eqref{eq:Prop6b} be denoted by $d_{\rm{thr}}$. Then, it is obtained that
\begin{gather}\label{eq:distRelProp6_7}
\CRLB_{\ell_1,\ell_2}=R_{\ell_3}\left({K_{\ell_3}P_J}/{d_{\rm{thr}}^\nu}+{N_0}/{2}\right)\,.
\end{gather}
Since $\CRLB_{\ell_1,\ell_2}<\CRLB_{\ell_1,\ell_3}$ as stated in the proposition, the equalizer property in Corollary~1 implies that $\CRLB_{\ell_1,\ell_2}<\CRLB_{\ell_1}(\bs{z}_{\ell_1,\ell_3}^{\rm{opt}})$, which, via \eqref{eq:CRLBm} and \eqref{eq:distRelProp6_7}, leads to
\vspace{-0.1cm}\begin{gather}\label{eq:distRelProp6_8}
d_{\rm{thr}}>\|\bs{x}_{\ell_3}-\bs{z}_{\ell_1,\ell_3}^{\rm{opt}}\|
\geq\|\bs{x}_{\ell_3}-\bs{z}_2\|
\end{gather}
where the last inequality follows from the definition of $\bs{z}_2$. From \eqref{eq:distRelProp6_7} and \eqref{eq:distRelProp6_8}, it is obtained that $\CRLB_{\ell_3}(\bs{z}_2)>\CRLB_{\ell_1,\ell_2}$. Since $\min\{\CRLB_{\ell_1}(\bs{z}_2),\CRLB_{\ell_2}(\bs{z}_2),\CRLB_{\ell_3}(\bs{z}_2)\}$ is upper bounded by $\CRLB_{\ell_1,\ell_2}$ by definition (see \eqref{eq:optProb_3tar_NoConstr} and \eqref{eq:optProb_2tar_NoConstr}), it can be concluded from the relation $\CRLB_{\ell_3}(\bs{z}_2)>\CRLB_{\ell_1,\ell_2}$ that $\CRLB_{\ell_3}(\bs{z}_2)$ is not a minimum of $\{\CRLB_{\ell_1}(\bs{z}_2),\CRLB_{\ell_2}(\bs{z}_2),\CRLB_{\ell_3}(\bs{z}_2)\}$. Then, a new location $\bs{z}_2^\delta$ can be defined as in the first case, and it can be shown that $\bs{z}_2$ cannot be optimal (cf. \eqref{eq:distRelProp6_4}--\eqref{eq:distRelProp6_6}).

Based on similar arguments to those in the two cases above, it can be shown that no locations on the straight line between $\bs{x}_{\ell_2}$ and $\bs{x}_{\ell_3}$ can be optimal.

Next, suppose that $\bs{z}_3$ is an optimal location for the jammer node, which lies on the straight line segment between $\bs{x}_{\ell_1}$ and $\bs{z}_{\ell_1,\ell_2}^{\rm{opt}}$ (excluding $\bs{z}_{\ell_1,\ell_2}^{\rm{opt}}$), where $\bs{z}_{\ell_1,\ell_2}^{\rm{opt}}$ is the optimizer of \eqref{eq:optProb_2tar_NoConstr} for target nodes $\ell_1$ and $\ell_2$, which corresponds to $\CRLB_{\ell_1,\ell_2}$. Since $\|\bs{x}_{\ell_1}-\bs{z}_{\ell_1,\ell_2}^{\rm{opt}}\|
>\|\bs{x}_{\ell_1}-\bs{z}_3\|$, it is obtained that
\begin{gather}\label{eq:distRelProp6_9}
\CRLB_{\ell_1}(\bs{z}_3)>\CRLB_{\ell_1}(\bs{z}_{\ell_1,\ell_2}^{\rm{opt}})=\CRLB_{\ell_1,\ell_2}
\end{gather}
where the equality is due to Corollary~1. Based on similar arguments to those in the first two cases, \eqref{eq:distRelProp6_9} implies that $\CRLB_{\ell_3}(\bs{z}_3)$ is not a minimum of $\{\CRLB_{\ell_1}(\bs{z}_3),\CRLB_{\ell_2}(\bs{z}_3),\CRLB_{\ell_3}(\bs{z}_3)\}$. Then, a new location $\bs{z}_3^\delta$ can be defined as in the first case, and it can be shown that $\bs{z}_3$ cannot be optimal (cf. \eqref{eq:distRelProp6_4}--\eqref{eq:distRelProp6_6}).

Finally, if $\bs{z}_4$ is an optimal location for the jammer node, which lies on the straight line segment between $\bs{x}_{\ell_2}$ and $\bs{z}_{\ell_1,\ell_2}^{\rm{opt}}$ (excluding $\bs{z}_{\ell_1,\ell_2}^{\rm{opt}}$), it can be shown in a similar manner to the previous case that $\bs{z}_4$ cannot be optimal.

Overall, the only possible location on the boundary of the convex hull (triangle) is $\bs{z}_{\ell_1,\ell_2}^{\rm{opt}}$ for which $\CRLB_{\ell_1}(\bs{z}_{\ell_1,\ell_2}^{\rm{opt}})=\CRLB_{\ell_2}(\bs{z}_{\ell_1,\ell_2}^{\rm{opt}})$ due to Corollary~1. Hence, if the optimal jammer location is on the boundary of the triangle formed by target nodes $\ell_1$, $\ell_2$, and $\ell_3$, then the optimizer of \eqref{eq:optProb_3tar_NoConstr} is equal to $\bs{z}_{\ell_1,\ell_2}^{\rm{opt}}$.

{\textbf{Part b)}:} If the condition in \eqref{eq:Prop6b} holds, it then follows form \eqref{eq:CRLBm} that
$\CRLB_{\ell_3}(\bs{z}_{\ell_1,\ell_2}^{\rm{opt}})\geq \CRLB_{\ell_1,\ell_2}$.
Then, Proposition~2 can be invoked to conclude that $\bs{z}_{\ell_1,\ell_2}^{\rm{opt}}$ is the optimal jammer location corresponding to \eqref{eq:optProb_3tar_NoConstr}. Hence, the optimal location for the jammer node is on the boundary of the convex hull (triangle) formed by the target nodes. To prove the necessity of \eqref{eq:Prop6b}, suppose that the optimal jammer location is on the boundary of the triangle. Then, the proof of Part a) shows that the optimal location for the jammer node is $\bs{z}_{\ell_1,\ell_2}^{\rm{opt}}$, which achieves a CRLB denoted by $\CRLB_{\ell_1,\ell_2}$. Due to the formulation in \eqref{eq:optProb_3tar_NoConstr}, $\CRLB_{\ell_1,\ell_2}$ is equal to $\min\{\CRLB_{\ell_1}(\bs{z}_{\ell_1,\ell_2}^{\rm{opt}}),
\CRLB_{\ell_2}(\bs{z}_{\ell_1,\ell_2}^{\rm{opt}}),$
$\CRLB_{\ell_3}(\bs{z}_{\ell_1,\ell_2}^{\rm{opt}})\}$ in this scenario. Hence, $\CRLB_{\ell_3}(\bs{z}_{\ell_1,\ell_2}^{\rm{opt}})\geq \CRLB_{\ell_1,\ell_2}$ must hold, which, based on \eqref{eq:CRLBm}, leads to \eqref{eq:Prop6b}.\hfill$\blacksquare$


\subsection{{Proof of Proposition 7}}\label{app:Prop7}

Consider the optimal jammer placement problem in \eqref{eq:OptProblem2} in the absence of distance constraints:
\begin{gather}\label{eq:optProb_NoConstr}
\max_{\bs{z}}\,\min_{{m\in\{1,\ldots,N_T\}}}\,
\CRLB_m(\bs{z})
\end{gather}
where $\CRLB_m(\bs{z})$ is as in \eqref{eq:CRLBm}. The aim is to prove that the optimizer of \eqref{eq:optProb_NoConstr} and the corresponding optimal value are equal to $\bs{z}^{\rm{opt}}_{i,j,k}$ and $\CRLB_{i,j,k}$, respectively, which are as defined in the proposition. Based on Proposition~4, $\bs{z}^{\rm{opt}}_{i,j,k}$ lies on the convex hull (triangle in this case) formed by the locations of target nodes $i$, $j$, and $k$.

\underline{Case 1:} First, assume that $\bs{z}^{\rm{opt}}_{i,j,k}$ belongs to the \textit{interior} of the triangle formed by these target nodes. Then, from Proposition~5,
the max-min solution in \eqref{eq:optProb_3tar_NoConstr} for target nodes $i$, $j$, and $k$ equalizes the CRLBs of these target nodes; that is, $\CRLB_i(\bs{z}^{\rm{opt}}_{i,j,k})=\CRLB_j(\bs{z}^{\rm{opt}}_{i,j,k})=\CRLB_k(\bs{z}^{\rm{opt}}_{i,j,k})=\CRLB_{i,j,k}$. Next, consider target node $\ell^*$, which is different from target nodes $i$, $j$, and $k$.
Since all the targets are on the two dimensional space, $\bs{z}^{\rm{opt}}_{i,j,k}$ must be on one of the triangles formed by target node $\ell^*$ and any two of target nodes $i$, $j$, and $k$. Without loss of generality, let that triangle be formed by target nodes $\ell^*$, $i$ and
$j$ (see Fig.~\ref{fig:prop7both}), and let the max-min solution in \eqref{eq:optProb_3tar_NoConstr} for these three target nodes be denoted by $\CRLB_{i,j,\ell^*}$ with the corresponding optimizer of $\bs{z}^{\rm{opt}}_{i,j,\ell^*}$. Since $\CRLB_{i,j,\ell^*}\geq\CRLB_{i,j,k}$ by definition,
$\CRLB_{i,j,\ell^*}=\min\{\CRLB_i(\bs{z}^{\rm{opt}}_{i,j,\ell^*}),\CRLB_j(\bs{z}^{\rm{opt}}_{i,j,\ell^*}),\CRLB_{\ell^*}(\bs{z}^{\rm{opt}}_{i,j,\ell^*})\}
\geq\CRLB_{i,j,k}=\CRLB_i(\bs{z}^{\rm{opt}}_{i,j,k})=\CRLB_j(\bs{z}^{\rm{opt}}_{i,j,k})=\CRLB_k(\bs{z}^{\rm{opt}}_{i,j,k})$ must hold. Therefore, $\CRLB_i(\bs{z}^{\rm{opt}}_{i,j,\ell^*})\geq\CRLB_i(\bs{z}^{\rm{opt}}_{i,j,k})$ and $\CRLB_j(\bs{z}^{\rm{opt}}_{i,j,\ell^*})\geq\CRLB_j(\bs{z}^{\rm{opt}}_{i,j,k})$ are obtained, which imply that (cf. \eqref{eq:CRLBm})
\begin{gather}\label{eq:distRelations}
\|\bs{x}_i-\bs{z}^{\rm{opt}}_{i,j,\ell^*}\|\leq\|\bs{x}_i-\bs{z}^{\rm{opt}}_{i,j,k}\|,~
\|\bs{x}_j-\bs{z}^{\rm{opt}}_{i,j,\ell^*}\|\leq\|\bs{x}_j-\bs{z}^{\rm{opt}}_{i,j,k}\|.
\end{gather}
Next, consider the two possible cases for $\bs{z}^{\rm{opt}}_{i,j,k}$:

\begin{figure}
\vspace{-0.1cm}
\center
  \includegraphics[width=90mm]{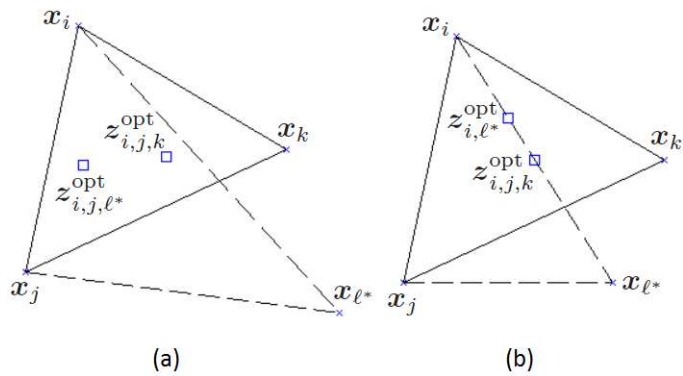}
  \vspace{-0.2cm}
\caption{Illustration for Case 1 of the proof of Proposition~7: (a) Case 1-(a), (b) Case 1-(b).}
 \label{fig:prop7both}
 \vspace{-0.3cm}
\end{figure}

\underline{Case 1-(a):} In this case, $\bs{z}^{\rm{opt}}_{i,j,k}$ belongs to the \textit{interior} of the triangle formed by target nodes $i$, $j$, and $\ell^*$, as shown in Fig.~\ref{fig:prop7both}-(a). Then, by Lemma~1 (see Appendix~\ref{app:ForProp5}), it follows from \eqref{eq:distRelations} that
\begin{gather}\label{eq:distRelation2}
\|\bs{x}_{\ell^*}-\bs{z}^{\rm{opt}}_{i,j,k}\|\leq\|\bs{x}_{\ell^*}-\bs{z}^{\rm{opt}}_{i,j,\ell^*}\|
\end{gather}
which implies $\CRLB_{\ell^*}(\bs{z}^{\rm{opt}}_{i,j,k})\geq\CRLB_{\ell^*}(\bs{z}^{\rm{opt}}_{i,j,\ell^*})$;
hence, the following relation is obtained:
\begin{align}\label{eq:l_star_larger1}
\CRLB_{\ell^*}(\bs{z}^{\rm{opt}}_{i,j,k})\geq\CRLB_{\ell^*}(\bs{z}^{\rm{opt}}_{i,j,\ell^*})
\geq\CRLB_{i,j,\ell^*}\geq\CRLB_{i,j,k}
\end{align}
where the second inequality follows from \eqref{eq:optProb_3tar_NoConstr} and the third inequality is due to the assumption in the proposition. The inequality in \eqref{eq:l_star_larger1} indicates that the optimal jammer location $\bs{z}^{\rm{opt}}_{i,j,k}$ obtained by considering target nodes $i$, $j$, and $k$ only results in a larger CRLB for target node $\ell^*$ than $\CRLB_{i,j,k}$, where $\ell^*$ is an arbitrary target node with $\ell^*\notin\{i,j,k\}$. Therefore, for the jammer node location $\bs{z}^{\rm{opt}}_{i,j,k}$, the objective function in \eqref{eq:optProb_NoConstr} becomes
\begin{gather}\label{eq:upperAchieved}
\min_{{m\in\{1,\ldots,N_T\}}}\,\CRLB_m(\bs{z}^{\rm{opt}}_{i,j,k})=\CRLB_{i,j,k}\,.
\end{gather}
Since $\CRLB_{i,j,k}$ is an upper bound on \eqref{eq:optProb_NoConstr} (since only three target nodes are considered in \eqref{eq:optProb_3tar_NoConstr}), which is achieved for $\bs{z}^{\rm{opt}}_{i,j,k}$ as specified in \eqref{eq:upperAchieved}, the solution of \eqref{eq:optProb_NoConstr} is given by $\bs{z}^{\rm{opt}}_{i,j,k}$ under the conditions in the proposition.

\underline{Case 1-(b):} In this case, $\bs{z}^{\rm{opt}}_{i,j,k}$ is on the edge of the triangle connecting target nodes $i$ and $\ell^*$, as shown in Fig.~\ref{fig:prop7both}-(b). (The same arguments below apply to the case in which $\bs{z}^{\rm{opt}}_{i,j,k}$ is on the edge of the triangle connecting target nodes $j$ and $\ell^*$.) Then, it is first obtained that $\CRLB_{i,\ell^*}\geq\CRLB_{i,j,\ell^*}\geq\CRLB_{i,j,k}$, where $\CRLB_{i,\ell^*}$ denotes the solution of \eqref{eq:optProb_2tar_NoConstr} for target nodes $i$ and $\ell^*$. Let $\bs{z}^{\rm{opt}}_{i,\ell^*}$ denote the optimizer of \eqref{eq:optProb_2tar_NoConstr} that results in $\CRLB_{i,\ell^*}$. Due to the equalizer solutions corresponding to $\CRLB_{i,\ell^*}$ and $\CRLB_{i,j,k}$ (see Corollary~1 and Proposition~5), $\CRLB_{i,\ell^*}\geq\CRLB_{i,j,k}$ implies that $\CRLB_i(\bs{z}^{\rm{opt}}_{i,\ell^*})\geq\CRLB_i(\bs{z}^{\rm{opt}}_{i,j,k})$. Hence,
the distance between $\bs{z}^{\rm{opt}}_{i,\ell^*}$ and target node $i$ is smaller than or equal to the distance between $\bs{z}^{\rm{opt}}_{i,j,k}$ and target node $i$. Since both $\bs{z}^{\rm{opt}}_{i,\ell^*}$ and $\bs{z}^{\rm{opt}}_{i,j,k}$ are on the straight line segment connecting target nodes $i$ and $\ell^*$, the following distance relation is obtained: $\|\bs{x}_{\ell^*}-\bs{z}^{\rm{opt}}_{i,j,k}\|\leq\|\bs{x}_{\ell^*}-\bs{z}^{\rm{opt}}_{i,\ell^*}\|$, which leads to $\CRLB_{\ell^*}(\bs{z}^{\rm{opt}}_{i,j,k})\geq\CRLB_{\ell^*}(\bs{z}^{\rm{opt}}_{i,\ell^*})$; hence, it follows that
\begin{align}\label{eq:l_star_larger2}
\CRLB_{\ell^*}(\bs{z}^{\rm{opt}}_{i,j,k})
\geq\CRLB_{\ell^*}(\bs{z}^{\rm{opt}}_{i,\ell^*})
\geq\CRLB_{i,\ell^*}
\geq\CRLB_{i,j,\ell^*}
\geq\CRLB_{i,j,k}\,.
\end{align}
Then, arguments similar to those in Case~1-(a) can be employed to prove that the solution of \eqref{eq:optProb_NoConstr} is given by $\bs{z}^{\rm{opt}}_{i,j,k}$ in this case, as well.


\underline{Case 2:} Secondly, consider the case in which $\bs{z}^{\rm{opt}}_{i,j,k}$ is on the \textit{boundary} of the triangle formed by target nodes $i$, $j$, and $k$. Let $\bs{z}^{\rm{opt}}_{i,j,k}$ be on the straight line connecting target nodes $i$ and $j$ without loss of generality.
Then, from Proposition~6, the jammer location $\bs{z}^{\rm{opt}}_{i,j,k}$ equalizes the CRLBs for target nodes $i$ and $j$, and is given by the optimal solution of \eqref{eq:optProb_2tar_NoConstr} corresponding to target nodes $i$ and $j$; that is, $\bs{z}^{\rm{opt}}_{i,j,k}=\bs{z}^{\rm{opt}}_{i,j}$ and $\CRLB_{i,j,k}=\CRLB_{i,j}$. Since the network consisting of the target nodes $i$ and
$j$ is a subnetwork of the network consisting of the target nodes $i$, $j$, and $\ell^*$, the following relation holds: $\CRLB_{i,j,\ell^*}\leq\CRLB_{i,j}$. On the other hand, since $\CRLB_{i,j,k}\leq\CRLB_{i,j,\ell^*}$ by definition, $\CRLB_{i,j}\leq\CRLB_{i,j,\ell^*}$ must also hold. Therefore, $\CRLB_{i,j}=\CRLB_{i,j,\ell^*}$ in this case, and it can shown that $\bs{z}^{\rm{opt}}_{i,j}=\bs{z}^{\rm{opt}}_{\ell^*,i,j}$ is the only possibility. Then, based on similar arguments to those in Case~1, it can be shown that target node $\ell^*$ has no effect on the optimal solution for all $\ell^*\notin\{i,j,k\}$; i.e., the solution of \eqref{eq:optProb_NoConstr} is given by $\bs{z}^{\rm{opt}}_{i,j,k}=\bs{z}^{\rm{opt}}_{i,j}$ under the conditions in the proposition.\hfill$\blacksquare$


%

\bibliographystyle{IEEEtran}

\bibliography{Ref4}

\footnotesize

${}$

\parbox[l]{1.1in}{\includegraphics[width=0.12\textwidth]{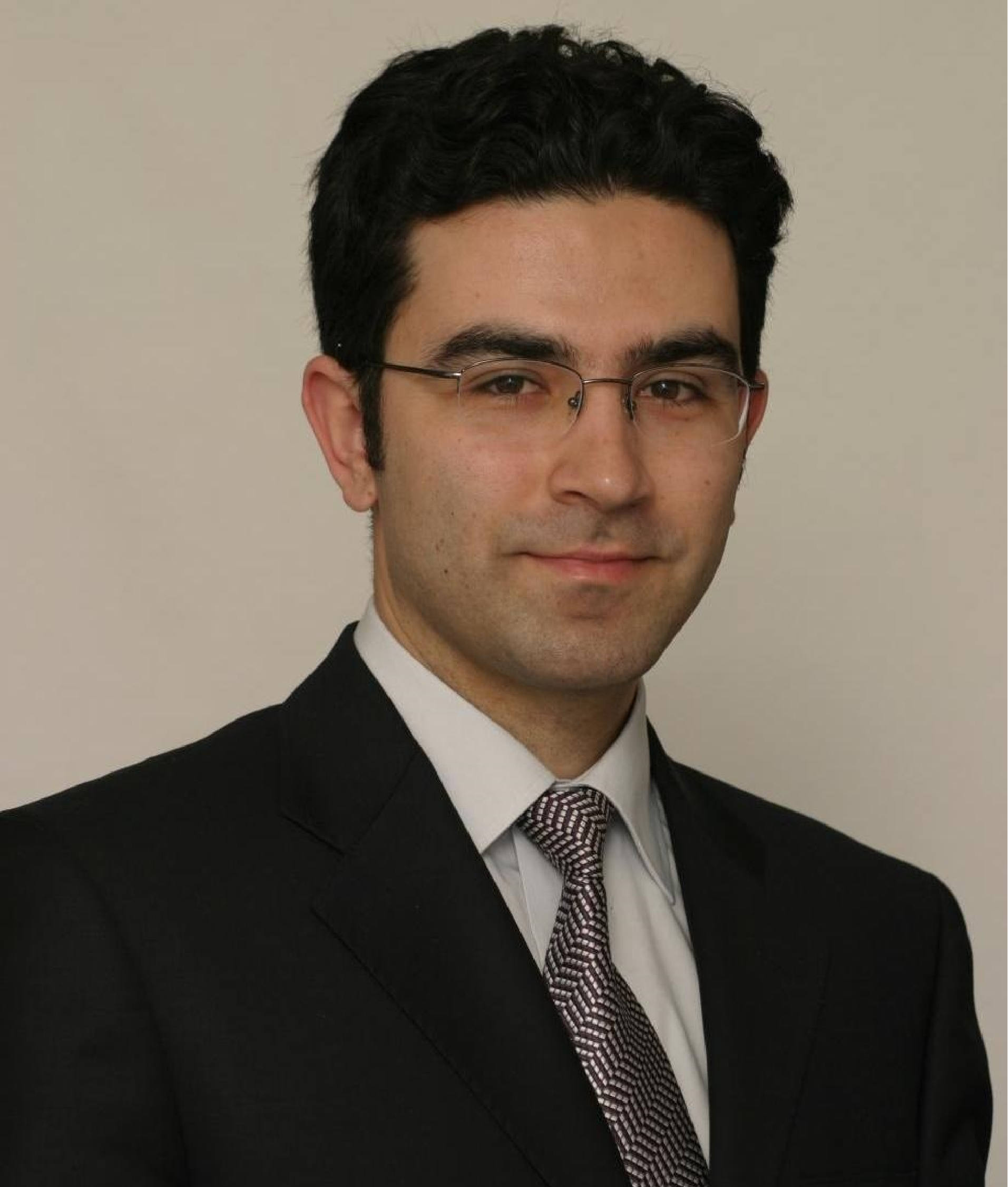}}
\parbox[l]{5.85in}{\textbf{Sinan  Gezici} (S'03--M'06--SM'11) received the B.S. degree from Bilkent University, Turkey in 2001, and the Ph.D. degree in Electrical Engineering from Princeton University in 2006. From 2006 to 2007, he worked at Mitsubishi Electric Research Laboratories, Cambridge, MA. Since 2007, he has been with the Department of Electrical and Electronics Engineering at Bilkent University, where he is currently an Associate Professor. Dr. Gezici’s research interests are in the areas of detection and estimation theory, wireless communications, and localization systems. Among his publications in these areas is the book Ultra-wideband Positioning Systems: Theoretical Limits, Ranging Algorithms, and Protocols (Cambridge University Press, 2008). Dr. Gezici is an associate editor for IEEE Transactions on Communications, IEEE Wireless Communications Letters, and Journal of Communications and Networks.}

${}$

${}$

\parbox[l]{1.1in}{\includegraphics[width=0.12\textwidth]{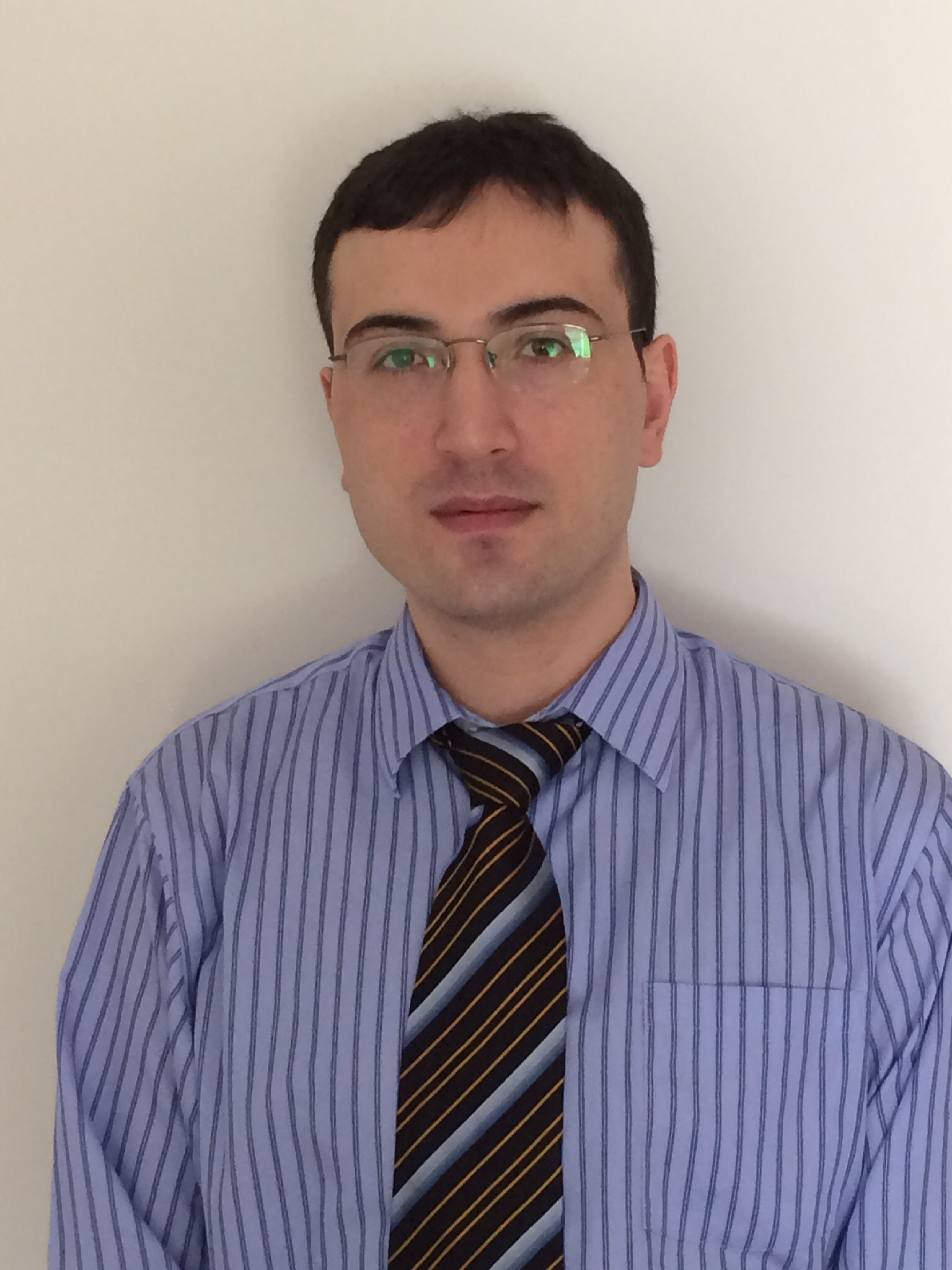}}
\parbox[l]{5.85in}{\textbf{Suat Bayram} received the B.S. degree from Middle East Technical University, Ankara, Turkey in 2007, and the M.S. and the Ph.D. degrees from Bilkent University, Ankara, Turkey, in 2009 and 2011, respectively. He is currently an Associate Professor in the Department of Electrical and Electronics Engineering at Turgut Ozal University, where he has been a faculty member since 2013. His research interests are in the fields of statistical signal processing and communications.}

${}$

\parbox[l]{1.1in}{\includegraphics[width=0.12\textwidth]{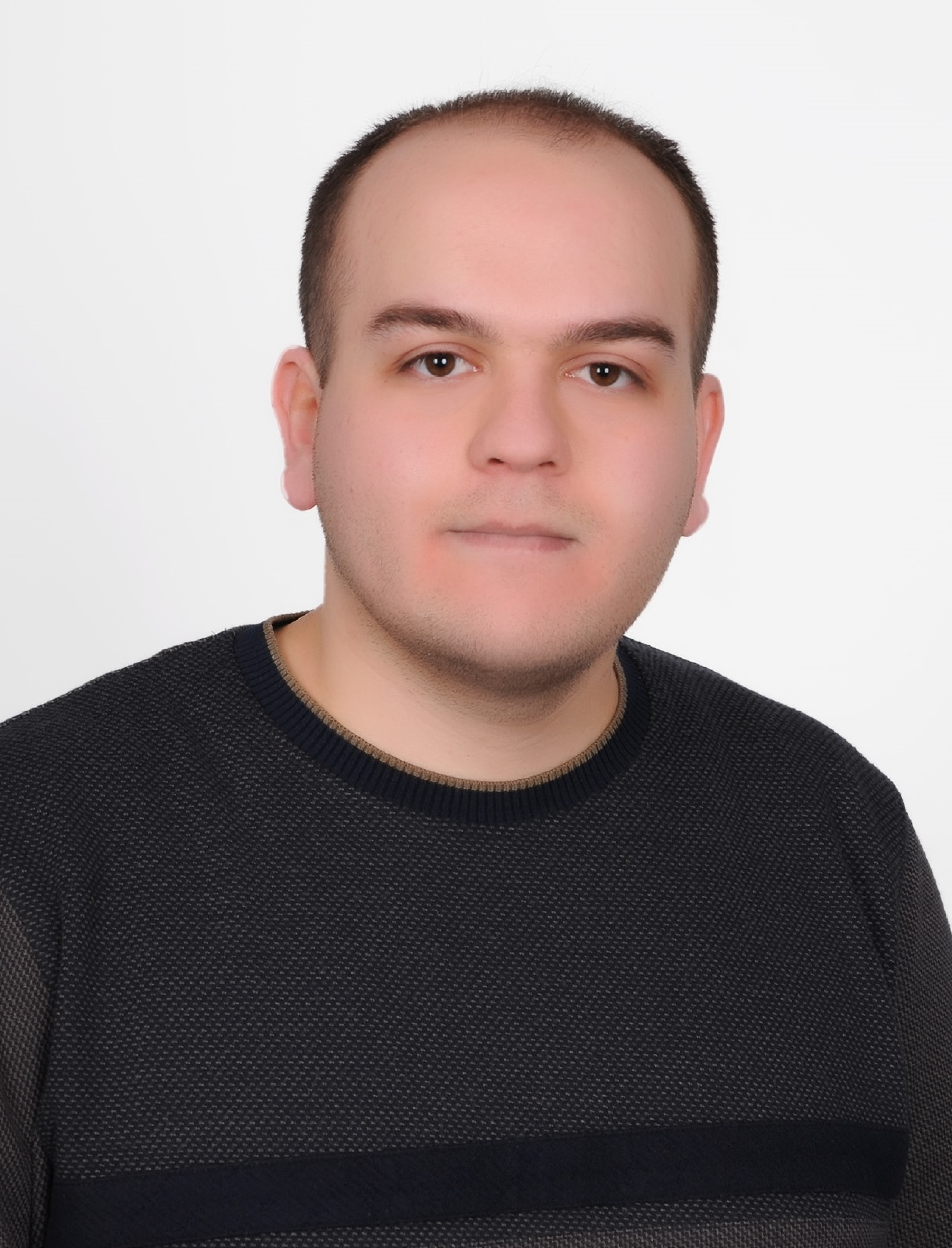}}
\parbox[l]{5.85in}{\textbf{Mehmet Necip Kurt} received the B.S. degree in 2014 from the Department of Electrical and Electronics Engineering at Bilkent University. Currently, he is working towards the M.S. degree in the same department. His main research interests are in the fields of wireless communications, detection and estimation theory, and wireless localization.}

${}$

\parbox[l]{1.1in}{\includegraphics[width=0.12\textwidth]{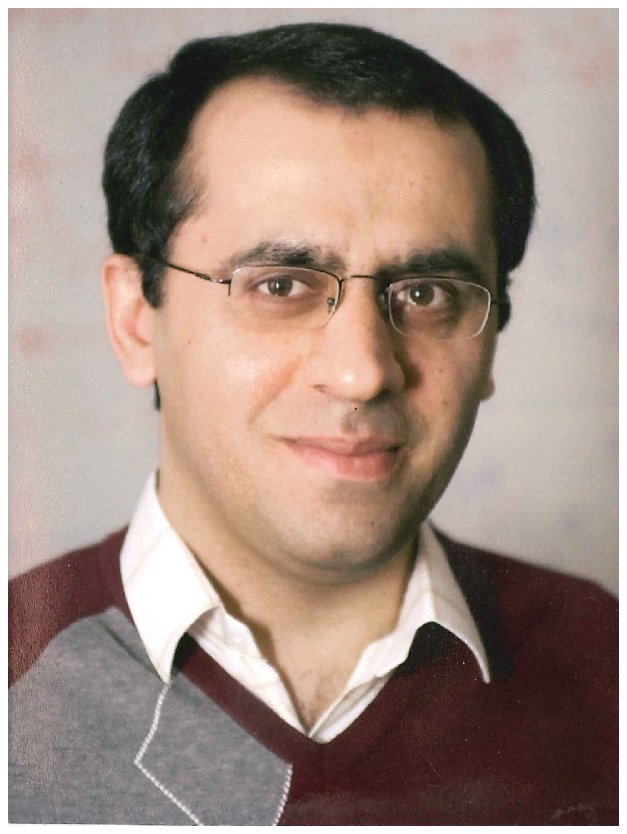}}
\parbox[l]{5.85in}{\textbf{Mohammad Reza Gholami} (S'09--M'14) received the Ph.D. degree from Chalmers University of Technology, Gothenberg, Sweden, in 2013. From 2014 to 2015, he was a postdoc research fellow at the department of signal processing, KTH, Stockholm, Sweden. From December 2012 to August 2013, he was a visiting researcher at the Adaptive Systems Laboratory, University of California, Los Angeles (UCLA).  His main interests are statistical inference, distributed estimation, wireless network positioning, network wide synchronization, and  data analytics.}

\end{document}